\documentclass[journal]{IEEEtran}
\usepackage{graphicx,amssymb,amsmath,color,textcomp,bm,cite,stfloats,algorithm,algorithmic,array,multirow,makecell,booktabs,subfigure,bbm,stmaryrd,amsmath,diagbox}
\usepackage[hidelinks]{hyperref}
\allowdisplaybreaks[1]

\begin{document}

\title{ RIS- and Multi-Snapshot-Enabled SISO 3D Position and Velocity Estimation With Single Base Station 

\thanks{This work was supported by the National Natural Science Foundation of China (Grant No.~62371044). {\it (Corresponding author: Yuyao Shen.)}   }

\thanks{ Yirun Wang, Yongqing Wang, and Yuyao Shen are with the School of Information and Electronics, Beijing
Institute of Technology, Beijing 100081, China   
(e-mail: yrwang719@163.com; wangyongqing@bit.edu.cn; syyxyz@gmail.com).  }
\thanks{ Gongpu Wang is with the Engineering Research Center of Network Management Technology for High Speed Railway of Ministry of Education, School of Computer Science and Technology, Beijing Jiaotong University, Beijing 100044, China (e-mail: gpwang@bjtu.edu.cn). }
\thanks{ Chintha Tellambura is with the Department of Electrical and Computer
Engineering, University of Alberta,  Edmonton, AB T6G 2R3, Canada 
(e-mail: ct4@ualberta.ca). }


\author{Yirun Wang, Yongqing Wang, Yuyao Shen, Gongpu Wang, and Chintha Tellambura, {\it  Fellow,  IEEE}}

}


\maketitle

\begin{abstract}
Reconfigurable intelligent surface (RIS) panels can act as cost-effective anchors for radio localization, complementing conventional base station (BS) anchors. This paper investigates joint three-dimensional position and velocity estimation (3D-JPVE) in single-input single-output (SISO) systems with only one BS available. We first theoretically show that 3D-JPVE is infeasible when relying solely on a single RIS or on multiple snapshots alone. To address this, we propose combining RIS deployment with multi-snapshot utilization to enable realizable 3D-JPVE. A two-stage method is developed for multi-snapshot channel parameter estimation, comprising a tensor-based coarse estimation step followed by a maximum likelihood refinement step. In particular, we introduce a third-order tensor formulation to decompose the challenging 3D joint angle-of-departure and Doppler shift estimation (3D-JADE) into two tractable subproblems, which are jointly solved via a low-complexity alternating optimization approach. Building on the channel parameter estimates, we further design a two-stage low-complexity method for optimal 3D-JPVE: coarse estimation is obtained from differential measurements through linear equations, and the preliminary results are refined iteratively using the original measurements. Moreover, we derive the closed-form Cram\'er–Rao lower bound (CRLB) and show that the proposed 3D-JPVE method approaches CRLB-level accuracy. Simulation results confirm the statistical efficiency of the proposed estimators and demonstrate substantial 3D-JPVE performance gains when deploying active RIS compared to passive RIS.
	
\end{abstract}


\section{Introduction}


Reconfigurable intelligent surface (RIS) is emerging as a disruptive technology in the wireless domain and is widely regarded as a key enabler for 6G systems \cite{Liu2025ITU, Wu2019Intelligent, Liu2021Reconfigurable}.
A RIS consists of numerous elements that can dynamically control the phase, amplitude, and polarization of incident electromagnetic (EM) waves, thereby capable of reconfiguring the wireless propagation environment in a programmable manner \cite{Wu2021Intelligent}.
Extensive studies have shown that RIS can substantially improve communication system performance in terms of coverage \cite{Zeng2021Reconfigurable}, throughput \cite{Huang2022Transforming}, and energy efficiency \cite{Huang2019Reconfigurable}.
Beyond communications, RIS integration has also attracted increasing interest in radio localization \cite{Aghashahi2024Single, Wu2024Exploit, Li2024Variational, Rahal2024RISenabled, Huang2023Joint}.

In positioning systems, RIS typically functions as an artificial anchor with precisely known state information, complementing or even substituting conventional anchors such as base stations (BSs) \cite{Liu2024Leveraging, Chen2024MultiRISenabled}.
Compared to BSs, RIS anchors are highly cost-effective, as they require neither expensive radio-frequency (RF) components nor complex signal processing \cite{Wu2020Beamforming, Zhang2023Active}.
Their low profile and conformal geometry also enable flexible deployment on road infrastructure \cite{Palmucci2023Twotimescale}, building facades \cite{Yu2022Location}, or mobile platforms such as vehicles \cite{Li2024STARRIS} and drones \cite{Zheng2024JrCUP}, thereby supporting ubiquitous localization. Moreover, RISs enhance positioning accuracy through two complementary mechanisms.
First, deploying additional RISs introduces greater geometrical diversity \cite{Wang2024heterogeneous}.
Second, careful RIS profile design allows direct optimization of localization performance \cite{Zhang2021MetaLocalization}.

Apart from the positioning accuracy gains, RIS can also enable localization in sparse-BS scenarios \cite{Bjornson2022Reconfigurable}.
In \cite{Keykhosravi2021SISO}, the addition of a RIS as an extra anchor enabled localization and synchronization of a user equipment (UE) in a single-input single-output (SISO) system with only one BS.
This study was later extended to account for UE mobility and the spatial-wideband effect \cite{Keykhosravi2022RISenabled}.
A key limitation, however, is that each anchor provides only one Doppler shift per epoch, restricting single-epoch velocity estimation to two-dimensional (2D) synchronized settings in the forgoing scenario \cite{Xie2022Intelligent}.
Since neither the prior velocity information on specific dimensions nor strict synchronization can be always assumed in practice, this work addresses joint three-dimensional (3D) velocity and position estimation using a single BS and RIS under clock imperfections.

In two-step position and velocity determination frameworks, intermediate geometrical measurements (i.e., channel parameters) are first extracted from the received signal, forming the basis for subsequent state estimation \cite{Zheng2024JrCUP}. 
Recently, tensor-based channel estimation (CE) has thus attracted significant attention due to its strong decoupling ability \cite{Zhou2017Lowrank} and high estimation accuracy \cite{Gong2023ESPRITbased}.
Most existing tensor-based CE methods for RIS-assisted communication and localization, however, focus on static scenarios \cite{Lin2021Tensorbased, Zheng2022Compressed, Lin2022Channel, Xu2022Sparse}. 
In dynamic settings, the RIS-related Doppler shift and angle-of-departure (AOD) pair (azimuth and elevation) become inherently coupled in the time domain, introducing new challenges for tensor-based parameter acquisition.
To address this three-dimensional joint AOD and Doppler estimation (3D-JADE) problem, \cite{He2024Slowmoving} proposed a  2D exhaustive search to estimate parameters for RIS-aided multiple-input multiple-output (MIMO) systems after tensor decomposition.  However, the 2D search incurs a high computational cost for acceptable accuracy.
An alternative method with fourth-order tensor formulation was introduced in \cite{Cai20254D}, yet it assumes a one-dimensional (1D) RIS panel, reducing the problem to a simplified 2D-JADE problem with only a single AOD to estimate.
The works \cite{Zhang2022Tensor,Zhang2024Integrated} similarly only investigated the degraded 2D-JADE problem for MIMO systems forming uniform linear arrays (ULAs).

\begin{figure}[t]
	\centering
	\includegraphics[width=0.72\linewidth]{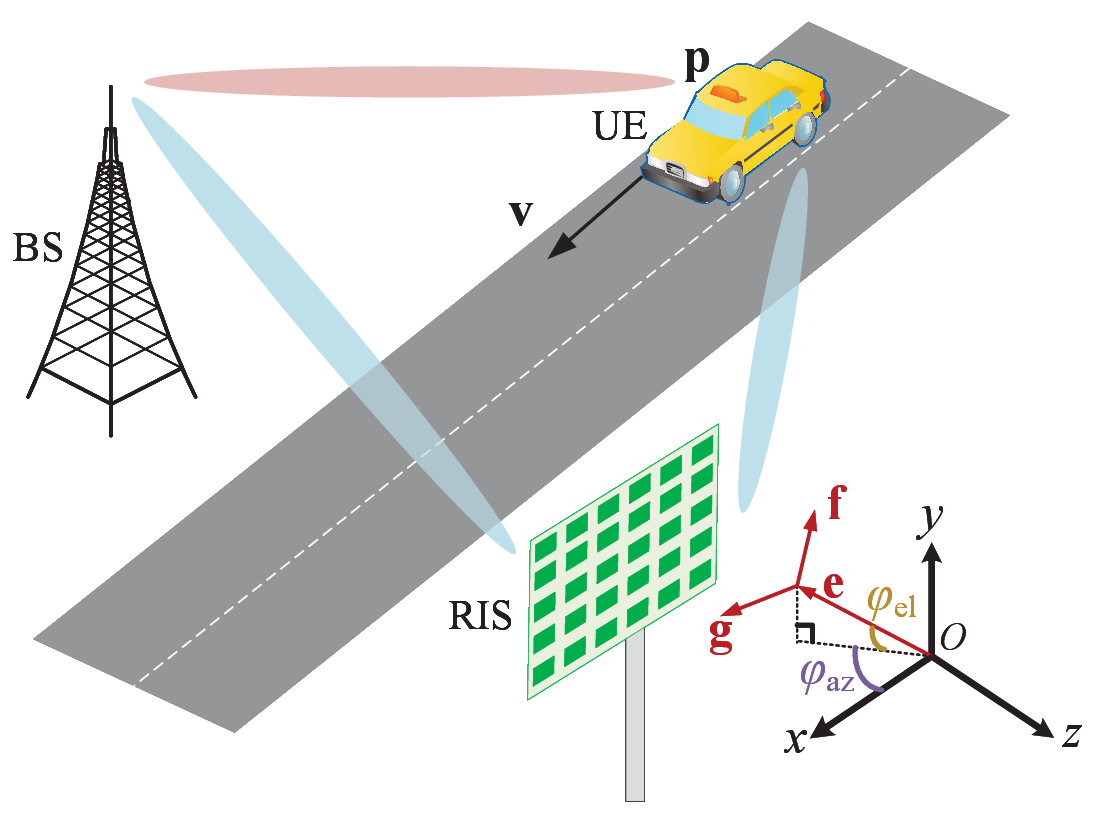}
	\caption{Illustration of RIS-empowered 3D-JPVE in the high-mobility scenario with a single BS.}
	\label{fig:system}
\end{figure}

Motivated by the above, we study in this paper the RIS- and multi-snapshot-enabled SISO joint 3D position and velocity estimation (3D-JPVE) with a single BS -- Fig.~\ref{fig:system}. 
We adopt the two-step estimation framework to achieve 3D-JPVE of a high-mobility, unsynchronized UE.   
A two-stage CE method is first proposed to extract multi-snapshot (i.e., multi-epoch) channel parameters.
Herein, the 3D-JADE issue is addressed via low-complexity tensor-based techniques in Stage~I, and the estimates are further refined using maximum likelihood estimation (MLE) in Stage~II. 
Next, another two-stage method is developed for 3D-JPVE using multi-epoch measurements.
The contributions of this work can be summarized as follows:
\begin{itemize}
	\item 
    We theoretically establish that 3D-JPVE is infeasible when relying solely on a single RIS or exclusively on multiple snapshots. To overcome these limitations, we propose integrating RIS deployment with multi-snapshot utilization, and we link multi-epoch measurements through state-transition models. 
    Within this framework, we show that 3D-JPVE under clock imperfections becomes feasible using as few as two snapshots. 
	\item 
	We propose a two-stage CE method to extract channel parameters from the received multi-snapshot signals. 
	Specifically, we design a tensor-based coarse estimation method, followed by a parameter refinement using MLE. 
	In the first stage, we develop a third-order tensor formulation scheme so that the challenging 3D-JADE is decomposed into two tractable 2D optimization issues with still total three variables. These two issues can be jointly solved via the proposed low-complexity alternating optimization approach, involving a decent initialization scheme and only one round of parameter refinement. 
	\item 
    We propose a two-stage low-complexity method for the optimal 3D-JPVE based on multi-epoch channel parameter estimates. In Stage~I, coarse state estimates are obtained using multi-epoch differential measurements, where the angle information provided by the RIS enables the reformulation of a linear estimation problem. In Stage~II, the original measurements are revisited, and a sequential error-compensation procedure is designed to refine the initial estimates iteratively.
	\item 
    We derive closed-form expressions for the Cram\'er-Rao lower bound (CRLB) on both channel parameter and UE state estimation. Theoretical analysis shows that the accuracy of the proposed 3D-JPVE method approaches the CRLB, and simulations further confirm the statistical efficiency of our estimators. Moreover, the results highlight the substantial accuracy gains by deploying active RIS over passive RIS for 3D-JPVE.
\end{itemize}

The remainder of this paper is organized as follows. 
Section~\ref{sec:PF} presents the geometry model and signal model. 
Section~\ref{sec:analysis} gives the feasibility analysis and CRLB derivation. 
Section~\ref{sec:CE} describes the proposed two-stage CE method and Section~\ref{sec:loc} introduces our two-stage 3D-JPVE method using channel parameters. 
Section~\ref{sec:NR} provides numerical results and finally Section~\ref{sec:conc} concludes this paper.

{\it Notations:} Scalars, vectors, matrices, and tensors are denoted by 
$ a $, $ \mathbf{a} $, $ \mathbf{A} $, and  $ \bm{\mathcal{A}} $, respectively. 
$ \mathbb{R} $ and $ \mathbb{C} $ denote the real and complex domains, respectively. 
Transpose, conjugate transpose, and inverse are denoted by $ (\cdot)^{\mathsf{T}} $, $ (\cdot)^{\mathsf{H}} $, and $ (\cdot)^{-1} $, respectively. 
Real part, imaginary part, and amplitude are denoted by $ \mathfrak{R}(\cdot) $, $ \mathfrak{I}(\cdot) $, and $ |\cdot| $, respectively, and $ \jmath^2=-1 $. 
$ \operatorname{diag}(\mathbf{a}) $ denotes the diagonal matrix formed by $ \mathbf{a} $, and $ \operatorname{blkdiag}(\mathbf{A}_1,\ldots,\mathbf{A}_L) $ is the block diagonal matrix formed by $ \mathbf{A}_1,\ldots,\mathbf{A}_L $. 
$ \|\cdot\| $, $ \|\cdot\|_{\mathsf{F}} $, $ \mathbb{E}[\cdot] $, and $ \operatorname{vec}(\cdot) $ denote the Euclidean norm, Frobenius norm, statistical expectation, and vectorization, respectively. 
Symbols $ \circ $, $ \otimes $, and $ \circledast $ denote the outer product, Kronecker product, and Hadamard product, respectively. 
$ \mathbf{1} $, $ \mathbf{0} $, and $ \mathbf{I} $ denote the all-one, all-zero, and identity matrices, respectively. 
Finally, $ \mathcal{N}(\bm{\mu},\bm{\Sigma}) $ and $ \mathcal{CN}(\bm{\mu},\bm{\Sigma}) $ denote real and complex Gaussian distributions with mean $ \bm{\mu} $ and variance $ \bm{\Sigma} $, respectively. 


\section{System Model} \label{sec:PF}

\subsection{System Setup}

This work considers deploying a RIS on the roadside to empower 3D-JPVE of a high-mobility, unsynchronized UE, as shown in Fig.~\ref{fig:system}.  
A single BS, with the known position $ \mathbf{q}_1\in\mathbb{R}^{3\times1} $, successively transmits orthogonal frequency-division multiplexing (OFDM) signal of $N$ frames. 
The UE receives the downlink signal of $ N $ snapshots at time instants $ t_1,\ldots,t_{N} $, respectively, and then acquires its position and velocity itself using signal processing techniques under clock imperfection condition. 
We consider the SISO system to investigate 3D-JPVE without any angle information available at the BS and UE ends. 

Since the timestamp of each snapshot differs, we assume the linear state-transition model for the UE's both position and clock bias during a short time period along its trajectory to connect the observations at different snapshots. 
Specifically, the position and clock bias of the UE at the $ n $th snapshot, denoted by $ \mathbf{p}_n\in\mathbb{R}^{3\times1} $ and $ B_n\in\mathbb{R},\, n=1,\ldots,N $, respectively, can be expressed as
\begin{align}
	\mathbf{p}_n &= \mathbf{p} + t_{n1} \mathbf{v} , \label{eq:pn} \\
	B_n &= B + t_{n1} D , \label{eq:Bn}
\end{align}
where $ \mathbf{p}\in\mathbb{R}^{3\times1} $ and $ B\in\mathbb{R} $ are the unknown initial position and clock bias, respectively, $ \mathbf{v}\in\mathbb{R}^{3\times1} $ and $ D\in\mathbb{R} $ are the unknown constant velocity and clock drift, respectively, and $ t_{n1} = t_n - t_1 $.
By assuming linear models \eqref{eq:pn}\textendash\eqref{eq:Bn}, the location and clock bias at all snapshots are both formulated with the initial value at the first snapshot and the constant variation rate.

The RIS forms a uniform planar array (UPA), located on the XOY plane in its local coordinate system (LCS) with $ M = M_{\mathrm{x}} \times M_{\mathrm{y}} $ elements. 
Other than the BS anchor, the RIS serves as another anchor with known position $ \mathbf{q}_2\in\mathbb{R}^{3\times1} $ and rotation matrix $ \mathbf{R}\in\textrm{SO}(3) $ (i.e., $ \mathbf{R} $ belongs to the special orthogonal group of 3D rotations \cite{Chen2022tutorial}). 
Like \cite{Zheng2024JrCUP,Zhang2023Active,Mylonopoulos2022Active}, this paper resorts to the active RIS enabling signal amplification via power supply to overcome the overwhelming power dominance of the direct BS-UE channel over the cascaded BS-RIS-UE channel. 
The active RIS can significantly enhance the 3D-JPVE performance compared to the conventional passive RIS (see Section~\ref{sec:RIS_comp}). 
The amplification coefficient for the incident signal is assumed to remain identical across active RIS elements, denoted by $ \eta>1 $. 
Notably, the active RIS introduces additional thermal noise when amplifying the signal, unlike the passive RIS \cite{Zhang2023Active}.

\subsection{Geometry Model}

This subsection describes the geometrical relationship among the system devices. 
We only concentrate on the line-of-sight (LOS) component of the direct BS-UE and cascaded BS-RIS-UE links for 3D-JPVE in this paper, as the multipath component (MPC) provides no extra information for position and velocity estimation in the SISO system \cite{Chen2024MultiRISenabled}. 
Denote the speed of light and the signal wavelength by $c$ and $\lambda$, respectively. 

Recap that the unsynchronized UE has clock bias $ B_n $ at the $n$th snapshot and clock drift $ D $.  
The delays over BS-UE and BS-RIS-UE links are given by
\begin{align}
	\tau_{\mathrm{L}}^{(n)} &= \| \mathbf{q}_1 - \mathbf{p}_n \|/c + B_n, \label{eq:tauL} \\
	\tau_{\mathrm{R}}^{(n)} &= \| \mathbf{q}_1 - \mathbf{q}_2 \|/c + \| \mathbf{q}_2 - \mathbf{p}_n \|/c + B_n. \label{eq:tauR}
\end{align}
Note that the delay over BS-RIS link, i.e., the first term in \eqref{eq:tauR}, is known. 
Let $ d_0 = \| \mathbf{q}_1 - \mathbf{q}_2 \| $ be the BS-RIS distance. 
Multiplying $ c $ on both sides of \eqref{eq:tauL}\textendash\eqref{eq:tauR} separately and subtracting known $d_0$ in \eqref{eq:tauR}, we have the pseudoranges $ d_{n,i} = \| \mathbf{q}_i - \mathbf{p}_n \| + B_n,\, i = 1,2 $ with $ B_n $ being in the unit of meter herein. 
Moreover, the Doppler shifts over BS-UE and BS-RIS-UE links are expressed as
\begin{align}
	\nu_{\mathrm{L}}^{(n)} &= \frac {\mathbf{v}^{\mathsf{T}} (\mathbf{q}_1 - \mathbf{p}_n)} 
	{\lambda \| \mathbf{q}_1 - \mathbf{p}_n \|} + D, \label{eq:nuL} \\
	\nu_{\mathrm{R}}^{(n)} &= \frac {\mathbf{v}^{\mathsf{T}} (\mathbf{q}_2 - \mathbf{p}_n)} 
	{\lambda \| \mathbf{q}_2 - \mathbf{p}_n \|} + D. \label{eq:nuR}
\end{align}
There is no Doppler shift in BS-RIS link due to the relatively stationary state. 
Multiplying $ \lambda $ on both sides of \eqref{eq:nuL}\textendash\eqref{eq:nuR} individually gives the pseudorange rates $ r_{n,i} = \mathbf{v}^{\mathsf{T}} (\mathbf{q}_i - \mathbf{p}_n)/\| \mathbf{q}_i - \mathbf{p}_n \| + D,\, i = 1,2 $ with $ D $ being in the unit of meter per second here.  

Define $ \bm{\varphi} = [\varphi_{\mathrm{\mathrm{az}}},\varphi_{\mathrm{\mathrm{el}}}]^{\mathsf{T}} $ containing a pair of azimuth and elevation angles. We have three unit-norm vectors that are orthogonal to each other, given by
\begin{align}
	\!\!\!\mathbf{e} (\bm{\varphi}) \!&=\! \big[ \cos (\varphi_{\mathrm{\mathrm{az}}}) \cos (\varphi_{\mathrm{\mathrm{el}}}), \sin (\varphi_{\mathrm{\mathrm{az}}}) \cos (\varphi_{\mathrm{\mathrm{el}}}), \sin (\varphi_{\mathrm{\mathrm{el}}}) \big]^{\mathsf{T}}, \label{eq:e} \\
	\!\!\!\mathbf{f} (\bm{\varphi}) \!&=\! [ -\sin (\varphi_{\mathrm{\mathrm{az}}}), \cos (\varphi_{\mathrm{\mathrm{az}}}), 0 ]^{\mathsf{T}}, \\
	\!\!\!\mathbf{g} (\bm{\varphi}) \!&=\! \big[ -\cos (\varphi_{\mathrm{\mathrm{az}}}) \sin (\varphi_{\mathrm{\mathrm{el}}}), -\sin (\varphi_{\mathrm{\mathrm{az}}}) \sin (\varphi_{\mathrm{\mathrm{el}}}), \cos (\varphi_{\mathrm{\mathrm{el}}}) \big]^{\mathsf{T}}\!. \label{eq:g}
\end{align}
Among these three vectors, the unit direction vector from the RIS to the UE can be expressed as
\begin{equation}
	\mathbf{e} (\bm{\varphi}_n) = \frac{ \mathbf{R}^{\mathsf{T}} (\mathbf{q}_2 - \mathbf{p}_n) }
	               { \| \mathbf{q}_2 - \mathbf{p}_n \| },
\end{equation}
where $ \bm{\varphi}_n = [\varphi_{n,\mathrm{az}},\varphi_{n,\mathrm{el}}]^{\mathsf{T}} $ is the AOD pair of RIS. 
Combining \eqref{eq:e}, the AODs of RIS can be obtained from
\begin{align}
		\varphi_{n,\mathrm{\mathrm{az}}} &= \operatorname{arctan2} \big( 
		\big[\mathbf{e} (\bm{\varphi}_n)\big]_2, 
		\big[\mathbf{e} (\bm{\varphi}_n)\big]_1 \big), \\
		\varphi_{n,\mathrm{\mathrm{el}}} &= \operatorname{arcsin} \big( 
		\big[\mathbf{e} (\bm{\varphi}_n)\big]_3 \big).
\end{align}
The angle-of-arrival (AOA) pair of RIS, say $ \bm{\varphi}_{\mathrm{A}} = [\varphi_{\mathrm{A},\mathrm{az}},\varphi_{\mathrm{A},\mathrm{el}}]^{\mathsf{T}} $, can be found in the same way as above. Note that $ \bm{\varphi}_{\mathrm{A}} $ is known and keep unchanged across different snapshots due to the fixed, known states of BS and RIS.


\subsection{Signal Model}

We consider the transmission of $G_0$ OFDM symbols in each frame $ n $, and assume that the first $G<G_0$ OFDM symbols contain pilots without loss of generality. 
The time interval between the $g$th OFDM symbol and the first one is denoted by $ T^{(g)} = (g-1)\Delta_{\mathrm{t}},\, g=1,\ldots,G $, where $ \Delta_{\mathrm{t}} $ is the duration of each OFDM symbol. 
Meanwhile, $ K_0 $ subcarriers are employed where the first  $ K<K_0 $ subcarriers are assumed to carry pilots. 
The frequency of the $k$th subcarrier is given by $ F^{(k)} = (k-1)\Delta_{\mathrm{f}},\, k=1,\ldots,K $, where $ \Delta_{\mathrm{f}}=B/K $ is the subcarrier spacing with the bandwidth $B$. 
This paper considers the identical pilot $x$ 
across distinct frames, OFDM symbols, and subcarriers. 

Collecting the received pilot signal at snapshot $n$ yields $ \mathbf{Y}\in\mathbb{C}^{K\times G} $, expressed as
\begin{equation}\label{eq:Y}
	\mathbf{Y}^{(n)} = \mathbf{Y}_{\mathrm{L}}^{(n)} + \mathbf{Y}_{\mathrm{R}}^{(n)} + \mathbf{N}^{(n)},
\end{equation}
where matrices $ \mathbf{Y}_{\mathrm{L}}^{(n)} $ and $ \mathbf{Y}_{\mathrm{R}}^{(n)} $ denote the received signal traveling through the BS-UE and BS-RIS-UE links, respectively, and $ \mathbf{N}^{(n)} $ is the noise matrix. 
These three terms on the right-hand side of \eqref{eq:Y} will be elaborated in the sequel.

Define $ \beta_{\mathrm{L}}^{(n)} = \alpha_{\mathrm{L}}^{(n)} x $ and $ \beta_{\mathrm{R}}^{(n)} = \alpha_{\mathrm{R}}^{(n)} x = \alpha_{\mathrm{R},1} \alpha_{\mathrm{R},2}^{(n)} x $, where $ \{ \alpha_{\mathrm{L}}^{(n)}, \alpha_{\mathrm{R}}^{(n)}, \alpha_{\mathrm{R},1}, \alpha_{\mathrm{R},2}^{(n)} \} $ are the complex path gains of BS-UE, BS-RIS-UE, BS-RIS, RIS-UE channels, respectively. The BS-RIS path gain $ \alpha_{\mathrm{R},1} $ can be measured in advance. 
Let 
$ \bm{\gamma}^{(n,g)} = \eta \big[  e^{\jmath\zeta_1^{(n,g)}}, \ldots, e^{\jmath\zeta_{M}^{(n,g)}} \big]^{\mathsf{T}}\in\mathbb{C}^{M\times1}  $ denote the active RIS profile. 
Define $ \mathbf{\Upsilon}^{(n)} = \big[\bm{\gamma}^{(n,1)},\ldots,\bm{\gamma}^{(n,G)}\big]^{\mathsf{T}} \in\mathbb{C}^{G\times M } $ as the RIS profile matrix over all OFDM symbols at snapshot $n$ and also define $ \bm{\Gamma}^{(n,g)} = \operatorname{diag}( \bm{\gamma}^{(n,g)} ) $.
Moreover, denote the local position matrix of RIS elements by 
$ \mathbf{P}_{\mathrm{R}} = \big[  
\mathbf{x}_{\mathrm{R}} \otimes \mathbf{1}_{M_{\mathrm{y}}\times1}, 
\mathbf{1}_{M_{\mathrm{x}}\times1} \otimes\mathbf{y}_{\mathrm{R}}, \mathbf{0}_{M\times1} \big] 
\in\mathbb{R}^{M\times 3} $ with every row representing the 3D relative position of a RIS element.  
Assume that $ \mathbf{x}_{\mathrm{R}} = [S^{(1)}, \ldots, S^{(M_{\mathrm{x}})} ]^{\mathsf{T}}\in\mathbb{R}^{M_{\mathrm{x}}\times1} $ and $ \mathbf{y}_{\mathrm{R}} = [S^{(1)}, \ldots, S^{(M_{\mathrm{y}})} ]^{\mathsf{T}}\in\mathbb{R}^{M_{\mathrm{y}}\times1} $, where $ S^{(m)} = (m-1) \Delta_{\mathrm{s}} $ with inter-element spacing $ \Delta_{\mathrm{s}} $.  
The RIS array response can be expressed as 
\begin{equation}\label{eq:a_org}
	\big[\mathbf{a}_{\mathrm{R}} (\bm{\varphi})\big]_m = 
	e^{\jmath \frac{2\pi}{\lambda} [\mathbf{P}_{\mathrm{R}}]_{m,:} \mathbf{e}(\bm{\varphi})},
	\; m=1,\ldots,M. 
\end{equation}
Define $ \breve{\mathbf{a}}_{\mathrm{R}} (\bm{\psi}^{(n)}_{\mathrm{R}}) = 
	\mathbf{a}_{\mathrm{R}} ( \bm{\varphi}_{\mathrm{A}} ) \circledast
	\mathbf{a}_{\mathrm{R}} ( \bm{\varphi}_{n} ) $, whose elements can be written as
\begin{equation}
	\!
	\big [\breve{\mathbf{a}}_{\mathrm{R}} (\bm{\psi}^{(n)}_{\mathrm{R}}) \big]_{ M_{\mathrm{y}}(m_{\mathrm{x}}-1) + m_{\mathrm{y}} } = 
	e^{ \jmath \frac{2\pi}{\lambda} ( S^{(m_{\mathrm{x}})} \psi_{n,{\mathrm{x}}} + S^{(m_{\mathrm{y}})} \psi_{n,{\mathrm{y}}} ) },\!
\end{equation}
where $ m_{\mathrm{x}} = 1,\ldots,M_{\mathrm{x}},\, m_{\mathrm{y}} = 1,\ldots,M_{\mathrm{y}}  $, and  $ \bm{\psi}^{(n)} = [ \psi_{n,\mathrm{x}}, \psi_{n,\mathrm{y}} ]^{\mathsf{T}} $ with 
\begin{align}
	\psi_{n,{\mathrm{x}}} &= 
	\cos( \varphi_{\mathrm{A},\mathrm{az}} ) \cos( \varphi_{\mathrm{A},\mathrm{el}} ) + \cos( \varphi_{n,\mathrm{az}} ) \cos( \varphi_{n,\mathrm{el}} ) ,
	\nonumber \\
	\psi_{n,{\mathrm{y}}} &= 
	\sin( \varphi_{\mathrm{A},\mathrm{az}} ) \cos( \varphi_{\mathrm{A},\mathrm{el}} ) + \sin( \varphi_{n,\mathrm{az}} ) \cos( \varphi_{n,\mathrm{el}} ).
\end{align}

Armed with the results above, the received signal matrices $ \mathbf{Y}_{\mathrm{L}}^{(n)} $ and $ \mathbf{Y}_{\mathrm{R}}^{(n)} $ can be written as
\begin{align}
	\mathbf{Y}_{\mathrm{L}}^{(n)} &= \beta_{\mathrm{L}}^{(n)} 
	\mathbf{F} \big(\tau_{\mathrm{L}}^{(n)}\big) \circledast \mathbf{T} \big(\nu_{\mathrm{L}}^{(n)}\big), \label{eq:YL} \\
	\mathbf{Y}_{\mathrm{R}}^{(n)} &= \beta_{\mathrm{R}}^{(n)} 
	\mathbf{F} \big(\tau_{\mathrm{R}}^{(n)}\big) \circledast \mathbf{T} \big(\nu_{\mathrm{R}}^{(n)}\big) \circledast \mathbf{S} \big(\bm{\psi}^{(n)}_{\mathrm{R}}\big), 
	\label{eq:YR}
\end{align}
where
\begin{align}
	\mathbf{F} (\tau) &= 
	\big[ 1, e^{-\jmath 2 \pi F^{(2)} \tau}, \ldots, e^{-\jmath 2 \pi F^{(K)} \tau} \big]^{\mathsf{T}} \mathbf{1}_{G\times1}^{\mathsf{T}}, \\
	\mathbf{T} (\nu) &= \mathbf{1}_{K\times1}
	\big[ 1, e^{\jmath 2\pi\nu T^{(2)}}, \ldots, e^{\jmath 2\pi\nu T^{(G)}} \big], \\
	\mathbf{S} (\bm{\psi}_{\mathrm{R}}^{(n)}) &= \mathbf{1}_{K\times1} \, \breve{\mathbf{a}}^{\mathsf{T}}_{\mathrm{R}} (\bm{\psi}_{\mathrm{R}}^{(n)})  (\bm{\Upsilon}^{(n)})^{\mathsf{T}}.
\end{align}

Furthermore, each element in the noise matrix $ \mathbf{N}^{(n)} $ is the sum of two parts, i.e., 
\begin{equation}\label{eq:Wngk}
	[\mathbf{N}^{(n)}]_{g,k} = n_{\mathrm{U}}^{(n,g,k)} + \bar{n}_{\mathrm{R}}^{(n,g,k)},
\end{equation}
with 
\begin{equation}
	\bar{n}_{\mathrm{R}}^{(n,g,k)} \!\!= \!
	\alpha_{\mathrm{R},2}^{(n)} 
	  e^{-\jmath 2 \pi F^{(k)} \tau_{\mathrm{R},2}^{(n)}}  e^{ \jmath 2 \pi \nu_{\mathrm{R}}^{(n)} T^{(g)} }\!\!
	\mathbf{a}_{\mathrm{R}}^{\mathsf{T}} ( \bm{\varphi}_n ) \bm{\Gamma}^{(n,g)}  \mathbf{n}_{\mathrm{R}}^{(n,g,k)},
\end{equation}
where $ \tau_{\mathrm{R},2}^{(n)} $ is the RIS-UE delay, $ n_{\mathrm{U}}^{(n,g,k)} $ denotes the i.i.d. thermal noise at UE with $ n_{\mathrm{U}}^{(n,g,k)} \sim \mathcal{CN}(0,\sigma_{\mathrm{U}}^2) $, and $ \mathbf{n}_{\mathrm{R}}^{(n,g,k)} $ is the i.i.d. thermal noise at the active RIS with $ \mathbf{n}_{\mathrm{R}}^{(n,g,k)} \sim \mathcal{CN}(\mathbf{0}_{M\times1},\sigma_{\mathrm{R}}^2\mathbf{I}_{M}) $.
It is worth noting that the second noise term in \eqref{eq:Wngk} is neglected in the passive RIS scenario. 
We can readily have such that $ [\mathbf{N}^{(n)}]_{g,k} \sim \mathcal{CN}(0, \sigma^2_n) $ with $ \sigma^2_n = \sigma_{\mathrm{U}}^2 + M\eta^2|\alpha_{\mathrm{R},2}^{(n)}|^2\sigma_{\mathrm{R}}^2 $ as noises $ n_{\mathrm{U}}^{(n,g,k)} $ and $ \mathbf{n}_{\mathrm{R}}^{(n,g,k)} $ are uncorrelated to each other. 

\section{Feasibility and CRLB Analysis}\label{sec:analysis}

At each snapshot, the received pilot signal can be leveraged to estimate the channel parameters for 3D-JPVE. The available multi-epoch parameters comprise: (i) pseudoranges $ \{ d_{n,1},d_{n,2} \}_{n=1}^N $, (ii) pseudorange rates $ \{ r_{n,1},r_{n,2} \}_{n=1}^N $, and (iii) angles $ \{ \varphi_{n,\mathrm{az}},\varphi_{n,\mathrm{el}} \}_{n=1}^N $. 
The aim is to estimate the initial 3D position $ \mathbf{p} $ and the constant 3D velocity $ \mathbf{v} $ with the existence of the initial clock bias $ B $ and the constant clock drift $ D $.

\subsection{Feasibility Analysis of 3D-JPVE}\label{sec:feasible}

Before discussing our RIS- and multi-snapshot-enabled 3D-JPVE scheme, we elaborate on two infeasible schemes at first. 

{\it Scheme A (Using RIS yet with single snapshot):}  In this scheme (i.e., $ N = 1 $), we have enough location-related parameters $ \{ d_{1,1},d_{1,2},\varphi_{1,\mathrm{az}},\varphi_{1,\mathrm{az}} \} $ to acquire the 3D position with the existence of clock bias \cite{Keykhosravi2021SISO, Keykhosravi2022RISenabled}. 
However, there are only two parameters $ \{ r_{1,1},r_{1,2} \} $ that are relevant to the velocity. 
As a result of the information deficiency (i.e, underdetermined equations), the estimation of 3D velocity is infeasible in {\it Scheme A}  even with the perfect frequency synchronization, not to mention the impact of clock drift. 

{\it Scheme B (Using multiple snapshots yet without RIS):} In this scheme, the multi-epoch JPVE-related parameters include $ \{ d_{n,1},r_{n,1} \}_{n=1}^N $. 
It seems that 3D-JPVE may be possible by accumulating parameters above across sufficient  snapshots, i.e., $ N\ge3 $ in the synchronized case and $ N\ge4 $ in the unsynchronized case. However, we demonstrate in Appendix~\ref{sec:proof} that {\it Scheme B} is still infeasible even with enough epochs, regardless of the synchronization state. 

{\it Proposed Scheme (Using RIS and multiple snapshots):} In our scheme, the available multi-epoch JPVE-related parameters are $ \{ d_{n,1},d_{n,2},r_{n,1},r_{n,2},\varphi_{n,\mathrm{az}},\varphi_{n,\mathrm{el}} \}_{n=1}^N $. By collecting parameters of at least two snapshots ($ N\ge2 $), the 3D-JPVE issue can be addressed somehow, regardless of the clock state. 
A straightforward solution is to leverage {\it Scheme A} to individually determine the 3D position at each epoch, and the 3D velocity can be obtained via the inter-epoch difference of the position. Nonetheless, the performance of this method is inferior due to error propagation in the sequential estimation process and also the neglect of information involved in pseudorange rates. The following Sections~\ref{sec:CE} and \ref{sec:loc} will introduce our proposed efficient 3D-JPVE framework.  

\subsection{CRLB Analysis}\label{sec:CRLB}

\subsubsection{Channel Parameter Estimation}

Firstly, define $ \alpha_{\mathrm{L},\mathfrak{R}}^{(n)} =  \mathfrak{R}(\alpha_{\mathrm{L}}^{(n)}) $, $ \alpha_{\mathrm{L},\mathfrak{I}}^{(n)} =  \mathfrak{I}(\alpha_{\mathrm{L}}^{(n)}) $, $ \alpha_{\mathrm{R},\mathfrak{R}}^{(n)} =  \mathfrak{R}(\alpha_{\mathrm{R},2}^{(n)}) $, and $ \alpha_{\mathrm{R},\mathfrak{I}}^{(n)} =  \mathfrak{I}(\alpha_{\mathrm{R},2}^{(n)}) $. 
Further define the JPVE-related channel parameter vector, overall channel parameter vector, the clock-irrelevant UE state vector, and the overal UE state vector as 
\begin{align}
	\!\bm{\eta} &= [ \ldots,d_{n,1},d_{n,2},r_{n,1},r_{n,2},\varphi_{n,\mathrm{az}},\varphi_{n,\mathrm{el}},\ldots ]^{\mathsf{T}}\in\mathbb{R}^{6N\times 1}, \nonumber \\
	\!\bm{\eta}_{\mathrm{ch}} & =  [ \ldots[\bm{\eta}]_{(6n-5):6n}, \alpha_{\mathrm{L},\mathfrak{R}}^{(n)}, \alpha_{\mathrm{L},\mathfrak{I}}^{(n)}, \alpha_{\mathrm{R},\mathfrak{R}}^{(n)}, \alpha_{\mathrm{R},\mathfrak{I}}^{(n)}\ldots ]^{\mathsf{T}} \!\!\!\in\!\mathbb{R}^{10N\times 1},
    \nonumber  \\
    \!\bm{\theta} &= [\mathbf{p}^{\mathsf{T}},\mathbf{v}^{\mathsf{T}}]^{\mathsf{T}} \in\mathbb{R}^{6\times 1},
	\nonumber  \\
	\!\bm{\xi} &= [\mathbf{p}^{\mathsf{T}},\mathbf{v}^{\mathsf{T}}, B, D]^{\mathsf{T}} \in\mathbb{R}^{8\times 1}.\label{eq:eta+xi}
\end{align}

The signal model in \eqref{eq:Y} can be reformed into a real-valued model in vector form as
\begin{equation}
	\tilde{\mathbf{y}} = \tilde{\mathbf{f}}(\bm{\eta}_{\mathrm{ch}}) + \tilde{\mathbf{n}},
\end{equation}
where $ \tilde{\mathbf{y}} = [ \ldots, \mathfrak{R}^{\mathsf{T}} (\mathbf{y}^{(n)}), \mathfrak{I}^{\mathsf{T}} (\mathbf{y}^{(n)}),\ldots ]^{\mathsf{T}} \in\mathbb{R}^{2NGK\times1} $ with $ \mathbf{y}^{(n)} = \operatorname{vec}(\mathbf{Y}^{(n)}) $, and $ \tilde{\mathbf{n}}\sim\mathcal{N}(\mathbf{0}_{2NGK\times1}, \bm{\Sigma}_{\mathrm{n}}) $ with the covariance matrix 
$ \bm{\Sigma}_{\mathrm{n}} = \operatorname{blkdiag} ( \tilde{\sigma}_1^2\mathbf{I}_{2GK}, \ldots, \tilde{\sigma}^2_N\mathbf{I}_{2GK} ) \in\mathbb{R}^{2NGK\times2NGK} $ and $ \tilde{\sigma}_n^2 = \sigma_n^2/2 $.  
Then the Fisher information matrix (FIM) with respect to $ \bm{\eta}_{\mathrm{ch}} $ is given by
\begin{equation}
	\bm{\Omega} (\bm{\eta}_{\mathrm{ch}}) = 
	\mathbf{J}_{\mathrm{ch}}^{\mathsf{T}} 
	\bm{\Sigma}_{\mathrm{n}}^{-1}
	\mathbf{J}_{\mathrm{ch}},
\end{equation}
where $ \mathbf{J}_{\mathrm{ch}} = \partial \tilde{\mathbf{f}}(\bm{\eta}_{\mathrm{ch}}) / \partial \bm{\eta}_{\mathrm{ch}}^{\mathsf{T}} \in \mathbb{R}^{2NGK \times 10N} $ is the Jacobian matrix and its elements are presented in Appendix~\ref{sec:Derivatives}.1.  
Note that $ \bm{\Omega} (\bm{\eta}_{\mathrm{ch}}) $ is a block diagonal matrix given by $ \bm{\Omega} (\bm{\eta}_{\mathrm{ch}}) = \operatorname{blkdiag}( \bm{\Omega}_1(\bm{\eta}_{\mathrm{ch}}),\ldots,\bm{\Omega}_N(\bm{\eta}_{\mathrm{ch}}) ) $ with $ \bm{\Omega}_n(\bm{\eta}_{\mathrm{ch}}) \in\mathbb{R}^{10\times10} $.  
Since each path gain is just a nuisance constant, we compute the equivalent FIM only with respect to JPVE-related channel parameters $ \bm{\eta} $ from 
\begin{equation}
	\!
    \bm{\Omega} (\bm{\eta}) = \operatorname{blkdiag}( \bm{\Omega}_1(\bm{\eta}),\ldots,\bm{\Omega}_N(\bm{\eta}) ),\;
    \bm{\Omega}_n(\bm{\eta}) \in\mathbb{R}^{6\times6},
\end{equation}
where \cite{Horn2012Matrix}
\begin{equation}
	\bm{\Omega}_n(\bm{\eta}) = 
	\bm{\Omega}_{n,1} - 
	\bm{\Omega}_{n,2} \bm{\Omega}_{n,3}^{-1} \bm{\Omega}_{n,2}^{\mathsf{T}},\;
	\bm{\Omega}_n(\bm{\eta}_{\mathrm{ch}}) =  \!\!
	\begin{array}{l}
		\setlength{\arraycolsep}{0.5pt}
		\left[ \begin{array}{cc}
			\bm{\Omega}_{n,1} & 
			\bm{\Omega}_{n,2}  \\ 
			\bm{\Omega}_{n,2}^{\mathsf{T}} & 
			\bm{\Omega}_{n,3}
		\end{array} \right]. 
	\end{array}
\end{equation}
Then the CRLB for the estimation of $ \bm{\eta} $ is
\begin{equation}
	\operatorname{CRLB} \left( [\bm{\eta}]_{\ell} \right) = 
	\big[ \bm{\Omega}^{-1}(\bm{\eta}) \big]_{\ell,\ell}, 
	\; \ell=1,\ldots,6N. 
\end{equation}

\subsubsection{UE State Estimation}

After channel parameter estimation at $N$ snapshots, the original measurement model (OMM) is given by
\begin{equation}\label{eq:model_loc}
	\hat{\bm{\eta}} = \mathbf{h} (\bm{\xi}) + \bm{\varsigma},
\end{equation}
where $ \hat{\bm{\eta}} $ and $ \bm{\varsigma} $ are the estimate and estimation error of $ \bm{\eta} $, respectively. 
The FIM with respect to $ \bm{\xi} $ is obtained using the chain rule of the FIM transformation \cite{Kay1993Fundamentals} as
\begin{equation}\label{eq:FIM0}
	\bm{\Omega} (\bm{\xi}) = 
	\mathbf{J}^{\mathsf{T}} 
	\bm{\Omega}  (\bm{\eta})
	\mathbf{J},
\end{equation}
where the covariance matrix of $ \bm{\varsigma} $ has been treated as $ \bm{\Omega} ^{-1} (\bm{\eta}) $, and the elements of Jacobian matrix $ \mathbf{J} = \partial \mathbf{h}(\bm{\xi}) / \partial \bm{\xi}^{\mathsf{T}} \in \mathbb{R}^{6N \times 8} $ are provided in Appendix~\ref{sec:Derivatives}.2. 
Consequently, the CRLB for the estimation of $ \bm{\xi} $ is
\begin{equation}
	\operatorname{CRLB} \left( [\bm{\xi}]_{\ell} \right) = 
	\big[ \bm{\Omega}^{-1}(\bm{\xi}) \big]_{\ell,\ell}, 
	\; \ell=1,\ldots,8. 
\end{equation}
The position error bound (PEB) and velocity error bound (VEB) are further defined as $ \sum_{\ell=1}^3 \operatorname{CRLB} \left( [\bm{\xi}]_{\ell} \right) $ and $ \sum_{\ell=4}^6 \operatorname{CRLB} \left( [\bm{\xi}]_{\ell} \right) $, respectively. 

We can also construct the differential measurement model (DMM)
\begin{equation}\label{eq:DMM}
	\hat{\bm{\eta}}_{\mathrm{d}} = \mathbf{h}_{\mathrm{d}} (\bm{\theta}) + \bm{\varsigma}_{\mathrm{d}},
\end{equation}
where 
$ \hat{\bm{\eta}}_{\mathrm{d}} $ and $ \bm{\varsigma}_{\mathrm{d}} $ are the estimate and estimation error of $ \mathbf{h}_{\mathrm{d}} (\bm{\theta}) $, respectively, with
\begin{equation}
	\mathbf{h}_{\mathrm{d}} (\bm{\theta}) = [ \ldots d_{n,1}-d_{n,2},r_{n,1}-r_{n,2},\varphi_{n,\mathrm{az}},\varphi_{n,\mathrm{el}} \ldots ]^{\mathsf{T}}\!\!\in\mathbb{R}^{4N\times 1}.
\end{equation}
The FIM with respect to $ \bm{\theta} $ is defined as
\begin{equation}\label{eq:FIMd}
	\bm{\Omega} (\bm{\theta}) = 
	\mathbf{J}_{\mathrm{d}}^{\mathsf{T}} 
	\bm{\Sigma}_{\mathrm{d}}^{-1}  (\bm{\eta})
	\mathbf{J}_{\mathrm{d}},
\end{equation}
where the Jacobian matrix $ \mathbf{J}_{\mathrm{d}} = \partial \mathbf{h}_{\mathrm{d}}(\bm{\theta}) / \partial \bm{\theta}^{\mathsf{T}} \in \mathbb{R}^{4N \times 6} $ can be easily obtained from $ \mathbf{J} $, and the covariance matrix $ \bm{\Sigma}_{\mathrm{d}} (\bm{\eta}) = \mathbb{E} [ \bm{\varsigma}_{\mathrm{d}} \bm{\varsigma}_{\mathrm{d}}^{\mathsf{T}} ] $ will be shown in Section~\ref{sec:loc_coarse}. The CRLB for $ \bm{\theta} $ is
\begin{equation}
	\operatorname{CRLB} \left( [\bm{\theta}]_{\ell} \right) = 
	\big[ \bm{\Omega}^{-1}(\bm{\theta}) \big]_{\ell,\ell}, 
	\; \ell=1,\ldots,6. 
\end{equation}

\section{Proposed Optimal Channel-Parameter Estimation Method}\label{sec:CE} 
This section illustrates our channel parameter acquisition method using a representative snapshot $n$. For clarity, the snapshot index $n$ is omitted from the subsequent notation. At each snapshot, the received pilot signal is individually processed by our CE algorithm to extract the corresponding multi-snapshot parameters. The framework first applies tensor-based estimators to obtain coarse parameter estimates (Sections~\ref{sec:tensor} and \ref{sec:CE_coarse}), which are then refined using the MLE principle for the optimal precision (Section~\ref{sec:CE_fine}). To better highlight the rationale, the tensor-based estimators are introduced under a noiseless setting, while in practice, the algorithm operates on noisy received signals.

\subsection{Tensor Construction and Estimation of Factor Matrices}\label{sec:tensor}

Define 
\begin{equation}
	\mathbf{a}^{ (L) }(\omega) = \big[ 0, e^{\jmath\omega}, \ldots, e^{\jmath(L-1)\omega} \big]^{\mathsf{T}} \in \mathbb{C}^{L\times 1}
\end{equation}
as a column of a Vandermonde matrix with length $ L $ and angular frequency $ \omega $, 
and let $ \mathbf{A}^{ (L) }(\omega) = \operatorname{diag}  (\mathbf{a}^{ (L) }(\omega)) $ be the diagonal form. 

The received signal matrices in \eqref{eq:YL}\textendash\eqref{eq:YR} can be recast as
\begin{align}
	\label{eq:YL1} 
	\mathbf{Y}_{\mathrm{L}} = \beta_{\mathrm{L}}  \mathbf{a}^{ (K) }(\omega_{d,1})& \circ \mathbf{a}^{ (G) } (\omega_{r,1}), \\
	\mathbf{Y}_{\mathrm{R}} = \beta_{\mathrm{R}} \mathbf{a}^{ (K) }(\omega_{d,2}) & \circ 
	\big( \mathbf{A}^{ (G) }(\omega_{r,2}) \bm{\Upsilon}
	\nonumber \\
	\times&  \big( \mathbf{a}^{ (M_{\mathrm{x}}) }(\omega_{\bm{\varphi},\mathrm{x}}) \otimes \mathbf{a}^{ (M_{\mathrm{y}}) }(\omega_{\bm{\varphi},\mathrm{y}})   \big)  \big),\label{eq:YR1}
\end{align}
where the angular frequencies are defined as
\begin{align}
	\omega_{d,1} = - &\frac{2\pi}{c} \Delta_{\mathrm{f}} d_{1},\;
	\omega_{d,2} = - \frac{2\pi}{c} \Delta_{\mathrm{f}} (d_{0}+d_2),\;
	\nonumber \\
	\omega_{r,1} &= \frac{2\pi}{\lambda} \Delta_{\mathrm{t}} r_{1},\;
	\omega_{r,2} = \frac{2\pi}{\lambda} \Delta_{\mathrm{t}} r_{2},
	\nonumber \\
	\omega_{\bm{\varphi},\mathrm{x}} &= \frac{2\pi}{\lambda} \Delta_{\mathrm{s}} \psi_{\mathrm{x}},\;
	\omega_{\bm{\varphi},\mathrm{y}} = \frac{2\pi}{\lambda} \Delta_{\mathrm{s}} \psi_{\mathrm{y}}.
\end{align}
Observe from \eqref{eq:YR1} that the three angular frequencies $ \{ \omega_{r,2}, \omega_{\bm{\varphi},\mathrm{x}}, \omega_{\bm{\varphi},\mathrm{y}} \} $ inherently couple together in the second dimension of $ \mathbf{Y}_{\mathrm{R}} $ (i.e., time domain), making the related estimator design challenging. 
The 2D-JADE method in \cite{Zhang2022Tensor,Zhang2024Integrated} tailored for the degraded ULA scenario with only a single AOD is not applicable anymore here.   
To address our extended 3D-JADE issue with an unknown AOD pair, we propose two tricks to convert the 3D estimation issue above into two tractable 2D subproblems, yet with still overall three parameters to be estimated. 

First, we set that
\begin{equation}
	\bm{\Upsilon} = 
	\mathbf{\Psi}_{\mathrm{x}}^{\mathsf{H}} \otimes \mathbf{\Psi}_{\mathrm{y}}^{\mathsf{H}},
\end{equation}
where $ \mathbf{\Psi}_{\mathrm{x}} \in\mathbb{C}^{M_{\mathrm{x}}\times G_1} $ and $ \mathbf{\Psi}_{\mathrm{y}} \in\mathbb{C}^{M_{\mathrm{y}}\times G_2} $ with $ G = G_1G_2 $. 
Among these two matrices, $\mathbf{\Psi}_{\mathrm{x}}$ is a pre-designed Vandermonde matrix, which will be used later in Section~\ref{sec:AOE}.  
Second, we notice that
\begin{equation}
	\mathbf{a}^{ (G) }(\omega_{r,i}) = 
	\mathbf{a}^{ (G_1) }(\tilde{\omega}_{r,i}) \otimes \mathbf{a}^{ (G_2) }(\omega_{r,i}), \; i=1,2, 
\end{equation}
where we have let $ \tilde{\omega}_{r,i} =  G_2\omega_{r,i} $. 
With these two results, \eqref{eq:YL1}\textendash\eqref{eq:YR1} can be further expressed as
\begin{align}\label{eq:YL2}
	\mathbf{Y}_{\mathrm{L}} &= \beta_{\mathrm{L}} \mathbf{a}^{ (K) }(\omega_{d,1}) \circ \big( \mathbf{a}^{ (G_1) }(\tilde{\omega}_{r,1}) \otimes \mathbf{a}^{ (G_2) }(\omega_{r,1})\big),	
	\\ \label{eq:YR2}
	\mathbf{Y}_{\mathrm{R}} &= \beta_{\mathrm{R}} \mathbf{a}^{ (K) }(\omega_{d,2}) \circ  \big(
	\big( \mathbf{A}^{ (G_1) }(\tilde{\omega}_{r,2})  \bm{\Psi}^{\mathsf{H}}_{\mathrm{x}} \mathbf{a}^{ (M_{\mathrm{x}}) }  (\omega_{\bm{\varphi},\mathrm{x}}) \big)
	\nonumber \\
	&\quad\quad\quad\quad\quad\quad\;\,
	\otimes \big( \mathbf{A}^{ (G_2) }(\omega_{r,2}) \bm{\Psi}^{\mathsf{H}}_{\mathrm{y}} \mathbf{a}^{ (M_{\mathrm{y}}) } (\omega_{\bm{\varphi},\mathrm{y}}) \big) \big).
\end{align}
Reshape $ \mathbf{Y} = \mathbf{Y}_{\mathrm{L}} + \mathbf{Y}_{\mathrm{R}} $ into a third-order tensor $ \bm{\mathcal{Y}}  $, which can be given in the canonical polyadic decomposition (CPD) format by
\begin{equation}
	\bm{\mathcal{Y}} = \llbracket \bm{\beta}; \mathbf{U}_1, \mathbf{U}_2, \mathbf{U}_3 \rrbracket \in\mathbb{C}^{K\times G_1 \times G_2},
\end{equation}
where
\begin{align}
	\bm{\beta} &= [\beta_{\mathrm{L}}, \beta_{\mathrm{R}}]^{\mathsf{T}} \in \mathbb{C}^{R\times 1},
	\nonumber \\
	\mathbf{U}_1 &= \big[  \mathbf{a}^{ (K) }(\omega_{d,1}),\, \mathbf{a}^{ (K) }(\omega_{d,2}) \big] \in \mathbb{C}^{K\times R},
	\nonumber \\
	\mathbf{U}_2 &= \big[ \mathbf{a}^{ (G_1) }(\tilde{\omega}_{r,1}),\, \mathbf{A}^{ (G_1) }(\tilde{\omega}_{r,2})  \bm{\Psi}^{\mathsf{H}}_{\mathrm{x}} \mathbf{a}^{ (M_{\mathrm{x}}) } (\omega_{\bm{\varphi},\mathrm{x}}) \big] \in \mathbb{C}^{G_1\times R},
	\nonumber \\
	\mathbf{U}_3 &= \big[ \mathbf{a}^{ (G_2) }(\omega_{r,1}),\, \mathbf{A}^{ (G_2) }(\omega_{r,2})  \bm{\Psi}^{\mathsf{H}}_{\mathrm{y}} \mathbf{a}^{ (M_{\mathrm{y}}) } (\omega_{\bm{\varphi},\mathrm{y}}) \big] \in \mathbb{C}^{G_2\times R},
\end{align}
where $ R=2 $ is the tensor rank. 
As such, only $ \{ \omega_{r,2}, \omega_{\bm{\varphi},\mathrm{x}}\} $ couple in mode~2 and $ \{ \omega_{r,2}, \omega_{\bm{\varphi},\mathrm{y}}\} $ couple in mode~3. 

Next, we compute the CPD of the third-order tensor $\bm{\mathcal{Y}}$. 
Note that the factor matrix $ \mathbf{U}_1 $ is a Vandermonde matrix. 
Consequently, we propose to leverage the Vandermonde structured CPD (VSCPD) with spatial smoothing to obtain the coarse factor estimates merely via linear algebra \cite{Sorensen2013Blind}, which can be further refined using the alternating least squares (ALS) method \cite{Gong2023ESPRITbased}. 
As such, we can acquire the estimate of factor matrices, denoted by $ \{ \hat{\mathbf{U}}_1, \hat{\mathbf{U}}_2, \hat{\mathbf{U}}_3 \} $, up to scaling and permutation ambiguities (see \cite[Sec.~3.2]{Kolda2009Tensor} for their definitions).  
As long as we can form an extended fourth-order tensor after spatial smoothing (i.e., at least 2D in each mode), the CPD is readily unique, following \cite[Prop.~III.2]{Sorensen2013Blind}. 

\subsection{Coarse Estimation of Channel Parameters from Tensor Factor Estimates}\label{sec:CE_coarse}

Once the estimate of the angular frequencies is obtained, the channel parameters can be acquired from
\begin{gather}
	\hat{d}_{1} = - \frac{c}{2\pi\Delta_{\mathrm{f}}}\hat{\omega}_{d,1},\;
	\hat{d}_{2} = - \frac{c}{2\pi\Delta_{\mathrm{f}}}\hat{\omega}_{d,2} - d_0,
	\nonumber \\
	\hat{r}_{1} = \frac{\lambda}{2\pi\Delta_{\mathrm{t}}} \hat{\omega}_{r,1},\;
	\hat{r}_{2} = \frac{\lambda}{2\pi\Delta_{\mathrm{t}}} \hat{\omega}_{r,2},
	\nonumber \\
	\hat{\varphi}_{\mathrm{az}} = \operatorname{arctan2} \big( \hat{e}_{\mathrm{y}}, \hat{e}_{\mathrm{x}} \big),\;
	\hat{\varphi}_{\mathrm{el}} = \arccos \big( \sqrt{ \hat{e}_{\mathrm{x}}^2 + \hat{e}_{\mathrm{y}}^2 }  \big), \label{eq:para_est}
\end{gather}
where $ \hat{\omega} $ is the estimate of angular frequency $ \omega $, and we have let $  \hat{e}_{\mathrm{x}} = \lambda \hat{\omega}_{\bm{\varphi},\mathrm{x}} / (2\pi\Delta_{\mathrm{s}})  - \cos( \varphi_{\mathrm{A},\mathrm{az}} ) \cos( \varphi_{\mathrm{A},\mathrm{el}} ) $ and $  \hat{e}_{\mathrm{y}} =  \lambda \hat{\omega}_{\bm{\varphi},\mathrm{y}} / (2\pi\Delta_{\mathrm{s}}) - \sin( \varphi_{\mathrm{A},\mathrm{az}} ) \cos( \varphi_{\mathrm{A},\mathrm{el}} ) $. 
Therefore, we target the estimation of angular frequencies as follows. 
Let $ \hat{\mathbf{u}}_{\ell,1} $ and $ \hat{\mathbf{u}}_{\ell,2} $ denote the columns of $ \hat{\mathbf{U}}_{\ell},\,\ell=1,2,3 $ corresponding to the direct BS-UE and cascaded BS-RIS-UE links, respectively. 

\subsubsection{Estimation of $ \{\omega_{d,1}, \omega_{d,2}, \omega_{r,1}\} $}

Using the two columns of $ \mathbf{U}_1 $, estimates $ \hat{\omega}_{d,1} $ and $ \hat{\omega}_{d,2} $ can be obtained from the correlation-based estimator as
\begin{equation}\label{eq:omega_d}
	\hat{\omega}_{d,i} = \arg \max_{\omega_{d,i}}
	\frac{ | \hat{\mathbf{u}}_{1,i}^{\mathsf{H}} \mathbf{a}^{(K)}(\omega_{d,i}) | }{ \| \hat{\mathbf{u}}_{1,i} \| \| \mathbf{a}^{(K)}(\omega_{d,i}) \| },\; i=1,2.
\end{equation}
Per \cite[Appendix~A]{Zhou2017Lowrank}, the correlation-based estimator above is equivalent to MLE provided the estimation error of each element in factor estimates follows an i.i.d. complex Gaussian distribution. 
Moreover, this estimator overcomes the scaling ambiguity via normalization. 
The optimization in \eqref{eq:omega_d} can be addressed with low complexity by finding the roots of a polynomial equation \cite{Park2020Spatial} other than by the exhaustive 1D search. The other correlation-based optimization problems hereafter can also be solved by the root-finding approach, unless otherwise stated.  

In fact, we do not know which column of $ \hat{\mathbf{U}}_1 $ corresponds to each of the two links before \eqref{eq:omega_d} due to the permutation ambiguity. 
However, notice the result that the distance of the cascaded link is always larger than that of the direct link (see the triangle in Fig.~\ref{fig:system}), and hence we further have $ d_0 + d_2 > d_1 $ involving the clock bias. This finding can be utilized to resolve the permutation ambiguity after we compute the two pseudoranges using \eqref{eq:omega_d}. 

Concerning the estimation of $ \omega_{r,1} $, although both the two columns $ \hat{\mathbf{u}}_{2,1} $ and $ \hat{\mathbf{u}}_{3,1} $ contain the relevant information, we only leverage $ \hat{\mathbf{u}}_{2,1} $ to get the estimate.  
This is because that the scaling coefficient of $ \omega_{r,1} $  at the argument of complex numbers in mode~2 is larger than that in mode~3, leading to a higher estimation accuracy of $ \omega_{r,1} $ in mode~2 given the factor estimates of similar precision in these two modes (see Section~\ref{sec:CE_RMSE}).  
Therefore, the estimator of $ \omega_{r,1} $ can be constructed as
\begin{equation}\label{eq:omega_r1}
	\hat{\omega}_{r,1} = \arg\max_{\omega_{r,1}}
	\frac{ | \hat{\mathbf{u}}_{2,1}^{\mathsf{H}} \mathbf{a}^{(G_1)}(G_2\omega_{r,1}) | }{ \| \hat{\mathbf{u}}_{2,1} \| \| \mathbf{a}^{(G_1)}(G_2\omega_{r,1}) \| }.
\end{equation}

\subsubsection{\texorpdfstring{Estimation of $ \{\omega_{r,2}, \omega_{\bm{\varphi},\mathrm{x}}, \omega_{\bm{\varphi},\mathrm{y}}\} $ (3D-JADE)}{Estimation of $\{ \omega_{r,2}, \omega_{\varphi,\mathrm{x}}, \omega_{\varphi,\mathrm{y}} \}$}}\label{sec:AOE}

With $ \hat{\mathbf{u}}_{2,2} $, we first aim at formulating the vector $ \mathbf{\mathbf{a}}^{(M_{\mathrm{x}})} (\omega_{\bm{\varphi},\mathrm{x}} ) $ as a function of $ \omega_{r,2} $ to extract this angular frequency first (using information in mode~2 only again).  
One possible solution is to get the vector with given $ \omega_{r,2} $ using the least squares (LS) principle. However, this scheme requires $ G_1 \ge M_{\mathrm{x}} $ to meet the full-rank condition, posing tight restrictions on the quantity of pilots and also the measurement range without phase ambiguity. 
To develop estimators that also work when $ G_1 < M_{\mathrm{x}} $, we propose formulating the fundamental angular frequency $ \omega_{\bm{\varphi},\mathrm{x}} $ in terms of $ \omega_{r,2} $ using the transformed-space type of estimation of signal parameters via the rotational invariance technique (TS-ESPRIT). 
We briefly introduce the TS-ESPRIT method here to make this paper self-contained. 
TS-ESPRIT provides an algebraic solution of the generators $ \{e^{\jmath\omega_{\ell}}\}_{\ell} $ of an Vandermonde matrix $ \mathbf{A} $. The column space of $ \mathbf{A} $ is transformed into $ \mathbf{U} = \bm{\Psi}^{\mathsf{H}} \mathbf{A} $ by the transformation matrix $ \bm{\Psi} $, and only the estimated column space of $ \mathbf{U} $ is available.  
The pioneering works \cite{Wen2020Tensor,Wen2018Tensor} have shown that $ \mathbf{U} $ satisfies a shift-invariance property provided that $ \bm{\Psi} $ is Vandermonde constrained. 
With this property, each $ \omega_{\ell} $ can be acquired using only linear algebra.    

Recall that we have designed $ \bm{\Psi}_{\mathrm{x}} $ as a Vandermonde matrix, and the Vandermonde structure remains after right-multiplying $ \bm{\Psi}_{\mathrm{x}} $ by a diagonal matrix.  
Consequently, we can obtain $ \hat{\omega}_{\bm{\varphi},\mathrm{x}} (\omega_{r,2}) $ from $ \hat{\mathbf{u}}_{2,2} $ using TS-ESPRIT. 
Thus, the estimator of $ \omega_{r,2} $ can be formulated as
\begin{equation}\label{eq:hat_ome_r2}
	\hat{\omega}_{r,2} = \arg\max_{\omega_{r,2}}
	\frac{ \big| \hat{\mathbf{u}}_{2,2}^{\mathsf{H}} \mathbf{A}^{(G_1)}(G_2\omega_{r,2}) \bm{\Psi}_{\mathrm{x}}^{\mathsf{H}} \bar{\mathbf{a}}^{(M_{\mathrm{x}})}(\omega_{r,2}) \big| }
	{ 
	\big\| \mathbf{A}^{(G_1)}(G_2\omega_{r,2}) \bm{\Psi}_{\mathrm{x}}^{\mathsf{H}} \bar{\mathbf{a}}^{(M_{\mathrm{x}})}(\omega_{r,2}) \big\| }, 
\end{equation}
where we have defined $ \bar{\mathbf{a}}^{(M_{\mathrm{x}})}(\omega_{r,2}) = \mathbf{a}^{(M_{\mathrm{x}})}(\hat{\omega}_{\bm{\varphi},\mathrm{x}}(\omega_{r,2})) $, and the constant norm in the denominator has been removed. 
The optimization problem in \eqref{eq:hat_ome_r2} can be solved via a 1D search. 
With the assistance of $ \hat{\omega}_{r,2} $, $ \{ \hat{\omega}_{\bm{\varphi},\mathrm{x}} , \hat{\omega}_{\bm{\varphi},\mathrm{y}} \} $ can be obtained from
\begin{align}
	\hat{\omega}_{\bm{\varphi},\mathrm{x}} &= \arg\max_{\omega_{\bm{\varphi},\mathrm{x}}}
	\frac{ \big| \hat{\mathbf{u}}_{2,2}^{\mathsf{H}} \mathbf{A}^{(G_1)}(G_2\hat{\omega}_{r,2}) \bm{\Psi}_{\mathrm{x}}^{\mathsf{H}} \mathbf{a}^{(M_{\mathrm{x}})}(\omega_{\bm{\varphi},\mathrm{x}}) \big| }
	{ 
	\big\| \mathbf{A}^{(G_1)}(G_2\hat{\omega}_{r,2}) \bm{\Psi}_{\mathrm{x}}^{\mathsf{H}} \mathbf{a}^{(M_{\mathrm{x}})}(\omega_{\bm{\varphi},\mathrm{x}}) \big\| },
	\nonumber \\
	\hat{\omega}_{\bm{\varphi},\mathrm{y}} &= \arg\max_{\omega_{\bm{\varphi},\mathrm{y}}}
	\frac{ \big| \hat{\mathbf{u}}_{3,2}^{\mathsf{H}} \mathbf{A}^{(G_2)}(\hat{\omega}_{r,2}) \bm{\Psi}_{\mathrm{y}}^{\mathsf{H}} \mathbf{a}^{(M_{\mathrm{y}})}(\omega_{\bm{\varphi},\mathrm{y}}) \big| }
	{ 
	\big\| \mathbf{A}^{(G_2)}(\hat{\omega}_{r,2}) \bm{\Psi}_{\mathrm{y}}^{\mathsf{H}} \mathbf{a}^{(M_{\mathrm{y}})}(\omega_{\bm{\varphi},\mathrm{y}}) \big\| }. \label{eq:hat_omega_phi}
\end{align}

Although TS-ESPRIT poses no extra restrictions on the system parameter design, its estimation accuracy is a bit limited \cite{Zheng2024JrCUP}. To improve the estimation performance, we perform a round of alternating optimization as follows. With $  \hat{\omega}_{\bm{\varphi},\mathrm{x}} $, we can refine $ \hat{\omega}_{r,2} $ from
\begin{equation}\label{eq:omega_r2}
	\hat{\omega}_{r,2} = \arg\max_{\omega_{r,2}}
	\frac{ \big| \hat{\mathbf{u}}_{2,2}^{\mathsf{H}} \operatorname{diag}(\bm{\Psi}_{\mathrm{x}}^{\mathsf{H}} \mathbf{a}^{(M_{\mathrm{x}})}(\hat{\omega}_{\bm{\varphi},\mathrm{x}})) \mathbf{a}^{(G_1)}(G_2\omega_{r,2})  \big| }
	{ 
	\big\| \operatorname{diag}(\bm{\Psi}_{\mathrm{x}}^{\mathsf{H}} \mathbf{a}^{(M_{\mathrm{x}})}(\hat{\omega}_{\bm{\varphi},\mathrm{x}})) \mathbf{a}^{(G_1)}(G_2\omega_{r,2}) \big\| }.
\end{equation}
Then we can obtain new $ \hat{\omega}_{\bm{\varphi},\mathrm{x}} $ and $ \hat{\omega}_{\bm{\varphi},\mathrm{y}} $ using \eqref{eq:hat_omega_phi} again. 
Note that the procedure above can be extended iteratively. However, simulations indicate that our alternating optimization algorithm reaches convergence after the first round of refinement, and further iterations do not improve the estimation performance. 
Therefore, our estimator \eqref{eq:hat_ome_r2} acts as an excellent initialization scheme so that the subsequently refinement procedure can be streamlined. 

\subsection{Refined Estimation of Channel Parameters via MLE}\label{sec:CE_fine}

The coarse CE method provides high accuracy with low computational complexity by minimizing the need for searches and iterations. However, further refinement of the estimates remains possible, allowing optimal performance at the cost of increased complexity. 

Vectorizing each matrix in \eqref{eq:Y} yields 
\begin{equation}
	\mathbf{y} = \bm{\Xi}\, \bm{\alpha} + \mathbf{n},
\end{equation}
where $ \bm{\Xi} = [\bm{\mu}_{\mathrm{L}},\bm{\mu}_{\mathrm{R}}]\in\mathbb{C}^{GK\times 2} $ and $ \bm{\alpha} = [\alpha_{\mathrm{L}}, \alpha_{\mathrm{R},2}] ^{\mathsf{T}} $ with 
$ \bm{\mu}_{\mathrm{L}} = \operatorname{vec}(\mathbf{Y}_{\mathrm{L}}) / \alpha_{\mathrm{L}} $ and $ \bm{\mu}_{\mathrm{R}} = \operatorname{vec}(\mathbf{Y}_{\mathrm{R}}) / \alpha_{\mathrm{R},2} $. Since each noise follows an i.i.d. complex Gaussian distribution with variance $ \sigma^2 $, the MLE of $ \{ \bm{\eta}, \bm{\alpha} \} $ is equivalent to the LS estimator, i.e., 
\begin{equation}
	[\hat{\bm{\eta}}, \hat{\bm{\alpha}}] = \arg\min_{\bm{\eta},\bm{\alpha}}
	\| \mathbf{y} - \bm{\Xi} (\bm{\eta}) \bm{\alpha} \|^2.
\end{equation}
Since each path gain is a complex constant, we formulate the nuisance vector $ \bm{\alpha} $ as a function of $ \bm{\eta} $ by letting $ \partial \| \mathbf{y} - \bm{\Xi} \, \bm{\alpha} \|^2 / \partial \bm{\alpha}  = \mathbf{0}_{2\times 1}  $, which gives $ \hat{\bm{\alpha}} (\bm{\eta}) = (\bm{\Xi}^{\mathsf{H}} \bm{\Xi})^{-1} \bm{\Xi}^{\mathsf{H}} \mathbf{y} $. 
Then the MLE issue with respect to nuisance-free channel parameters can be expressed as
\begin{equation}\label{eq:ML}
	\hat{\bm{\eta}} = \arg\min_{\bm{\eta}}
	\big\| \mathbf{y} - \bm{\Xi} (\bm{\eta}) \big(\bm{\Xi}^{\mathsf{H}}(\bm{\eta}) \bm{\Xi}(\bm{\eta})\big)^{-1} \bm{\Xi}^{\mathsf{H}}(\bm{\eta}) \mathbf{y} \big\|^2.
\end{equation} 
The optimization problem above can be solved via, e.g., gradient descent algorithm, with $ \hat{\bm{\eta}} $ obtained from our coarse CE method as initialization.  

The proposed two-stage CE method is summarized in Algorithm~\ref{alg:CE}.
The procedure involves only three steps that require iteration or search, while the remaining components admit low-complexity algebraic solutions.
First, the ALS refinement for factor estimation has complexity $ \mathcal{O}(NI_1GK) $, where $I_1$ denotes the number of iterations.
Second, the correlation-based estimator \eqref{eq:hat_ome_r2} requires a  complexity of $ \mathcal{O}(NI_2 G_1M_{\mathrm{x}}) $ with $I_2$ searches. 
Here, one round of 1D search suffices, avoiding the higher-dimensional search used in \cite{He2024Slowmoving}. 
Third, the parameter refinement via MLE costs a complexity of $ \mathcal{O}(NI_3GK) $ with $I_3$ iterations. 

\begin{algorithm}[t]
	\caption{Two-Stage Channel Parameter Estimation.}
	\renewcommand{\algorithmicrequire}{\textbf{Input:}}
	\renewcommand{\algorithmicensure}{\textbf{Output:}}
	\label{alg:CE}
	\begin{algorithmic}[1]
		\REQUIRE 
		Received signal matrices $ \{\mathbf{Y}^{(n)}\}_n^N $ at $N$ snapshots.
		\renewcommand{\algorithmicrequire}{\textbf{Implementation:}}
		\REQUIRE
		\FOR{$n=1$ to $N$}
		\STATE Reshape $ \mathbf{Y}^{(n)} $ into the third-order tensor $ \bm{\mathcal{Y}}^{(n)} $.
		\STATE Utilize VSCPD and ALS techniques to get the factor estimates of $ \bm{\mathcal{Y}}^{(n)} $. 
		\STATE Compute $ \{ \hat{\omega}_{d,1}^{(n)}, \hat{\omega}_{d,2}^{(n)}, \hat{\omega}_{r,1}^{(n)} \} $ using \eqref{eq:omega_d}\textendash\eqref{eq:omega_r1}.
		\STATE Formulate $ \omega^{(n)}_{\bm{\varphi},\mathrm{x}} $ with $ \omega^{(n)}_{r,2} $ using TS-ESPRIT and compute $ \{ \hat{\omega}_{r,2}^{(n)}, \hat{\omega}_{\bm{\varphi},\mathrm{x}}^{(n)}, \hat{\omega}_{\bm{\varphi},\mathrm{y}}^{(n)} \} $ using \eqref{eq:hat_ome_r2}\textendash\eqref{eq:hat_omega_phi}.   
		\STATE Refine $ \{ \hat{\omega}_{r,2}^{(n)}, \hat{\omega}_{\bm{\varphi},\mathrm{x}}^{(n)}, \hat{\omega}_{\bm{\varphi},\mathrm{y}}^{(n)} \} $ using \eqref{eq:omega_r2} and \eqref{eq:hat_omega_phi}.
		\STATE Obtain coarse channel-parameter estimates using \eqref{eq:para_est}.  {\bf (End of Stage I)}
		\STATE Obtain fine estimates by solving MLE issue in \eqref{eq:ML} via iterations. {\bf (End of Stage II)}
		\ENDFOR  
		
		\ENSURE 
		Estimates  
		$ \{ \hat{d}_{n,1},\hat{d}_{n,2},\hat{r}_{n,1},\hat{r}_{n,2},\hat{\varphi}_{n,\mathrm{az}},\hat{\varphi}_{n,\mathrm{el}} \}_{n=1}^N $.
	\end{algorithmic}
\end{algorithm}  

\section{Proposed Optimal 3D-JPVE Method Using Channel Parameter Estimates}\label{sec:loc}

Channel parameter estimation at all $N$ snapshots yields 
\begin{gather}
	\hat{d}_{n,i} = d_{n,i} + \Delta d_{n,i},\; \hat{r}_{n,i} = r_{n,i} + \Delta r_{n,i},\;i=1,2,
	\nonumber \\
	\hat{\varphi}_{n,\mathrm{az}} = \varphi_{n,\mathrm{az}} + \Delta \varphi_{n,\mathrm{az}},\; 
	\hat{\varphi}_{n,\mathrm{el}} = \varphi_{n,\mathrm{el}} + \Delta \varphi_{n,\mathrm{el}}.
\end{gather}
Simulations show that these MLE-based channel parameter estimators attain the CRLB-level accuracy, and the estimation error of each parameter approximately follows a zero-mean Gaussian distribution.
Therefore, we assume that the estimation error vector $ \bm{\varsigma} $ in \eqref{eq:model_loc} satisfies $ \bm{\varsigma}\sim\mathcal{N}( \mathbf{0}_{6N\times 1}, \bm{\Sigma} (\bm{\eta}) ) $ with $ \bm{\Sigma} (\bm{\eta}) = \bm{\Omega}^{-1}(\bm{\eta}) $ in this section for ease of derivation.  
Nonetheless, we input the channel parameter estimates obtained earlier instead of ideal Gaussian random variables into our 3D-JPVE method to verify its robustness in Section~\ref{sec:sim_loc}. 

With $ \{ \hat{\varphi}_{n,\mathrm{az}}, \hat{\varphi}_{n,\mathrm{el}} \} _{n=1}^N $, we further use \eqref{eq:e}\textendash\eqref{eq:g} to construct  
\begin{equation}\label{eq:efg_hat}
	\hat{\mathbf{e}}_{n} = \mathbf{e}_n + \Delta \mathbf{e}_n,\;
	\hat{\mathbf{f}}_{n} = \mathbf{f}_n + \Delta \mathbf{f}_n,\;
	\hat{\mathbf{g}}_{n} = \mathbf{g}_n + \Delta \mathbf{g}_n.
\end{equation}
For notational simplicity, we replace $ \bm{\chi}(\bm{\varphi}_n) $ and $ \bm{\chi}(\hat{\bm{\varphi}}_n) $ with  $ \bm{\chi}_n $ and $ \hat{\bm{\chi}}_n,\,\bm{\chi}\in\{\mathbf{e}, \mathbf{f}, \mathbf{g}\} $, respectively, in \eqref{eq:efg_hat} and hereafter. 
Define $ d_{n,i}^{\circ} = d_{n,i} - B_n $ and $ r^{\circ}_{n,i} = r_{n,i} - D ,\, i = 1,2 $ as the true distances and distance rates. 

\subsection{Coarse Estimation Using Differential Measurement Model}\label{sec:loc_coarse}

In this subsection, we leverage the differential measurements and angle information to construct a linear equation set so that we can obtain a coarse estimate of position and velocity. 
We get differential measurements via 
\begin{align}\label{eq:dr12_hat}
	\hat{d}_{n,12} &= \hat{d}_{n,1} - \hat{d}_{n,2} = d_{n,12} + \Delta d_{n,12},
	\nonumber \\
	\hat{r}_{n,12} &= \hat{r}_{n,1} - \hat{r}_{n,2} = r_{n,12} + \Delta r_{n,12},
\end{align}
where 
\begin{gather}
	d_{n,12} = d_{n,1}^{\circ} - d_{n,2}^{\circ}, \;
	r_{n,12} = r_{n,1}^{\circ} - r_{n,2}^{\circ}, 
	\label{eq:dr12} \\
	\Delta d_{n,12} = \Delta d_{n,1} - \Delta d_{n,2},\; 
	\Delta r_{n,12} = \Delta r_{n,1} - \Delta r_{n,2}.
\end{gather}
In the following, we will first focus on the noise-free case to construct the linear equations, and next, we will further present the related noisy version. 

We start from the differential measurements in \eqref{eq:dr12}. 
In the first equation, moving $ d_{n,2}^{\circ} $ to the left-hand side and then squaring both sides, we obtain
\begin{equation}\label{eq:TDOA0}
	d_{n,12}^2 - \|\mathbf{q}_1\|^2 + \|\mathbf{q}_2\|^2 = 
	2 (\mathbf{q}_2^{\mathsf{T}} - \mathbf{q}_1^{\mathsf{T}} ) \mathbf{p}_n - 2 d_{n,12} \| \mathbf{q}_2 - \mathbf{p}_n \|
\end{equation}
with a tricky nuisance norm term $ \| \mathbf{q}_2 - \mathbf{p}_n \| $. Thanks to the angle information provided by the RIS, we have
\begin{equation}
	\| \mathbf{q}_2 - \mathbf{p}_n \| = 
	\mathbf{e}_n^{\mathsf{T}} \mathbf{e}_n \| \mathbf{q}_2 - \mathbf{p}_n \| 
	= \mathbf{e}_n^{\mathsf{T}} \mathbf{R}^{\mathsf{T}} (\mathbf{q}_2 - \mathbf{p}_n). 
\end{equation}
Then \eqref{eq:TDOA0} can be reformulated as
\begin{align}
	d_{n,12}^2 &+ 2 d_{n,12} \mathbf{e}_n^{\mathsf{T}} \mathbf{R}^{\mathsf{T}} \mathbf{q}_2 - \|\mathbf{q}_1\|^2 + \|\mathbf{q}_2\|^2  
	\nonumber \\ &=
	2( d_{n,12} \mathbf{e}_n^{\mathsf{T}} \mathbf{R}^{\mathsf{T}} - \mathbf{q}_1^{\mathsf{T}} + \mathbf{q}_2^{\mathsf{T}}  ) (\mathbf{p} + t_{n1} \mathbf{v}). \label{eq:TDOA}
\end{align}
Taking the time derivative on both sides of \eqref{eq:TDOA} yields
\begin{align}\label{eq:FDOA}
	&d_{n,12} r_{n,12} + r_{n,12} \mathbf{e}_n^{\mathsf{T}} \mathbf{R}^{\mathsf{T}} \mathbf{q}_2 
	\nonumber \\
	&=
	r_{n,12} \mathbf{e}_n^{\mathsf{T}} \mathbf{R}^{\mathsf{T}} \mathbf{p} + 
	( (t_{n1}r_{n,12} - d_{n,12}) \mathbf{e}_n^{\mathsf{T}}\mathbf{R}^{\mathsf{T}} + \mathbf{q}_1^{\mathsf{T}} - \mathbf{q}_2^{\mathsf{T}} ) \mathbf{v},
\end{align}
where we have used $ \partial d_{n,12}/ \partial t_{n1} = -r_{n,12}  $. 

Furthermore, recall that $ \{ \mathbf{e}_n, \mathbf{f}_n, \mathbf{g}_n \} $ are orthogonal to each other. 
The result $ \mathbf{f}_n^{\mathsf{T}} \mathbf{e}_n = 0 $ obtains
\begin{equation}\label{eq:AOA1}
	\mathbf{f}_n^{\mathsf{T}} \mathbf{R}^{\mathsf{T}} \mathbf{q}_2 = 
	\mathbf{f}_n^{\mathsf{T}} \mathbf{R}^{\mathsf{T}} \mathbf{p} + t_{n1} \mathbf{f}_n^{\mathsf{T}} \mathbf{R}^{\mathsf{T}} \mathbf{v}. 
\end{equation}
Analogously, owing to  $ \mathbf{g}_n^{\mathsf{T}} \mathbf{e}_n = 0 $, we also have
\begin{equation}\label{eq:AOA2}
	\mathbf{g}_n^{\mathsf{T}} \mathbf{R}^{\mathsf{T}} \mathbf{q}_2 = 
	\mathbf{g}_n^{\mathsf{T}} \mathbf{R}^{\mathsf{T}} \mathbf{p} + t_{n1} \mathbf{g}_n^{\mathsf{T}} \mathbf{R}^{\mathsf{T}} \mathbf{v}. 
\end{equation}

Collecting the results in \eqref{eq:TDOA}, \eqref{eq:FDOA}, \eqref{eq:AOA1}, and \eqref{eq:AOA2} across $ N $ epochs, we can construct a linear equation set 
$ \bm{\rho} = \bm{\Phi} \bm{\theta} $ with $ \bm{\rho}\in\mathbb{R}^{4N\times1} $ and $ \bm{\Phi}\in\mathbb{R}^{4N\times6} $.

Further replacing the true parameters in \eqref{eq:efg_hat}\textendash\eqref{eq:dr12_hat} by their noisy measurements, we construct 
\begin{equation}\label{eq:rho_hat}
	\hat{\bm{\rho}} = \hat{\bm{\Phi}} \bm{\theta} + \bm{\varepsilon}, 
\end{equation}
where $ \hat{\bm{\rho}} $ and $ \hat{\bm{\Phi}} $ are the noisy versions of $ \bm{\rho} $ and $ \bm{\Phi} $, respectively, and the elements of the new estimation error vector $ \bm{\varepsilon} $ are given in Appendix~\ref{sec:error}.  
Note that directly deriving the covariance matrix of $ \bm{\varepsilon} $ is challenging. 
To this end, we approximate $ \bm{\varepsilon}(\bm{\varsigma}_{\mathrm{d}}) $, with $\bm{\varsigma}_{\mathrm{d}}$ given in \eqref{eq:DMM}, using the first-order Taylor expansion at $ \bm{\varsigma}_{\mathrm{d}} = \mathbf{0}_{4N\times1}  $. That is, we have $ \bm{\varepsilon} \approx \mathbf{J}_{\mathrm{err}} (\bm{\theta}) \bm{\varsigma}_{\mathrm{d}} $. The Jacobian matrix $ \mathbf{J}_{\mathrm{err}} = \partial \bm{\varepsilon}(\bm{\varsigma}_{\mathrm{d}}) / \partial \bm{\varsigma}_{\mathrm{d}}^{\mathrm{T}} \in \mathbb{R}^{4N\times4N} $, a function of $ \bm{\theta} $, is shown in Appendix~\ref{sec:error}. 
Then we approximately obtain the covariance matrix of $ \bm{\varepsilon} $ as $ \mathbb{E} [\bm{\varepsilon}\bm{\varepsilon}^{\mathsf{T}}] = \mathbf{J}_{\mathrm{err}}(\bm{\theta}) \bm{\Sigma}_{\mathrm{d}}(\bm{\eta}) \mathbf{J}_{\mathrm{err}}^{\mathsf{T}}(\bm{\theta}) $. 
The elements of the covariance matrix $ \bm{\Sigma}_{\mathrm{d}} (\bm{\eta}) = \mathbb{E} [ \bm{\varsigma}_{\mathrm{d}} \bm{\varsigma}_{\mathrm{d}}^{\mathsf{T}} ] $ can be acquired using $ \bm{\Sigma}(\bm{\eta}) $. For instance, the variance of $ \Delta d_{n,12} $  and the covariance of $ \{ \Delta d_{n,12}, \Delta \varphi_{n,\mathrm{az}} \} $ satisfy
\begin{gather}
	\!\!\!\mathbb{E}[\Delta d_{n,12}^2] = \mathbb{E}[\Delta d_{n,1}^2] + \mathbb{E}[\Delta d_{n,2}^2] - 2 \mathbb{E}[\Delta d_{n,1}\Delta d_{n,2}],
	\nonumber \\
	\!\mathbb{E}[\Delta d_{n,12} \Delta \varphi_{n,\mathrm{az}}] \!=\! \mathbb{E}[\Delta d_{n,1} \Delta \varphi_{n,\mathrm{az}}] \!-\! \mathbb{E}[\Delta d_{n,2} \Delta \varphi_{n,\mathrm{az}}],\!\!
\end{gather}
and each term on the right-hand side can be found in $ \bm{\Sigma}(\bm{\eta}) $. 

With \eqref{eq:rho_hat}, the MLE of $ \bm{\theta} $ is equivalent to the following weighted least squares (WLS) estimation
\begin{equation}\label{eq:theta_hat}
	\hat{\bm{\theta}} = (\hat{\bm{\Phi}}^{\mathsf{T}} \mathbf{W}_{\mathrm{d}} \hat{\bm{\Phi}})^{-1} \hat{\bm{\Phi}}^{\mathsf{T}} \mathbf{W}_{\mathrm{d}} \hat{\bm{\rho}},
\end{equation}
where $ \mathbf{W}_{\mathrm{d}}(\bm{\eta},\bm{\theta}) = \mathbb{E}^{-1} [\bm{\varepsilon}\bm{\varepsilon}^{\mathsf{T}}] $ is the weight matrix, depending on $ \bm{\eta} $ and $ \bm{\theta} $. 
We point out that the estimate $ \hat{\bm{\eta}} $ has been obtained from CE, whereas $ \bm{\theta} $ is unknown at the beginning.  
To tackle this issue, we first set $ \mathbf{J}_{\mathrm{err}} \leftarrow \mathbf{I}_{4N} $ to construct $ \mathbf{W}_{\mathrm{d}}(\hat{\bm{\eta}}) $, and then we compute initial $ \hat{\bm{\theta}} $ using \eqref{eq:theta_hat}. 
Next, we update the weight matrix as $ \mathbf{W}_{\mathrm{d}}(\hat{\bm{\eta}}, \hat{\bm{\theta}}) $ and acquire new $ \hat{\bm{\theta}} $. Repeat the procedure above, and very few iterations  (i.e., typically two to four) suffice to converge. 

We analyze as follows the estimation error of $ \hat{\bm{\theta}} $, denoted by $ \Delta \hat{\bm{\theta}} = \hat{\bm{\theta}} - \bm{\theta}  $. 
With \eqref{eq:rho_hat} and \eqref{eq:theta_hat}, $ \Delta \hat{\bm{\theta}} $ is given by
\begin{equation}
	\Delta \hat{\bm{\theta}} = (\hat{\bm{\Phi}}^{\mathsf{T}} \mathbf{W}_{\mathrm{d}} \hat{\bm{\Phi}})^{-1} \hat{\bm{\Phi}}^{\mathsf{T}} \mathbf{W}_{\mathrm{d}} \bm{\varepsilon}. 
\end{equation}
Left-multiply \eqref{eq:DMM} by $ \mathbf{J}_{\mathrm{err}} $ and compare the result with \eqref{eq:rho_hat}, we have $ \mathbf{J}_{\mathrm{err}} \mathbf{h}_{\mathrm{d}}(\bm{\theta}) = \hat{\bm{\Phi}} \bm{\bm{\theta}} $. Further taking the derivative with respect to $ \bm{\theta}^{\mathsf{T}} $ on both sides obtains $ \hat{\bm{\Phi}} = \mathbf{J}_{\mathrm{err}} \mathbf{J}_{\mathrm{d}} $. With such result, the covariance matrix of $ \Delta \hat{\bm{\theta}} $ is $ ( \mathbf{J}_{\mathrm{d}}^{\mathsf{T}} \bm{\Sigma}_{\mathrm{d}}^{-1}(\bm{\eta}) \mathbf{J}_{\mathrm{d}} )^{-1} $, equal to the inverse of FIM in \eqref{eq:FIMd}.  
Thus, our estimator \eqref{eq:theta_hat} can attain the accuracy of CRLB under DMM. 

\subsection{Refined Estimation Using Original Measurement Model}

In the last subsection, to facilitate the construction of linear equations, the multi-epoch differential measurements were constructed. This, however, leads to information loss at the same time, and the previous estimation performance is suboptimal under OMM \cite{Hara2013perturbation}. 
To attain the optimality, we return to the OMM shown at \eqref{eq:model_loc} and utilize iterations to sequentially refine the estimates with the foregoing coarse result as reliable initialization. 

With $ \hat{\bm{\theta}} $ obtained earlier, it is straightforward to also acquire the estimate of clock parameters $ \{\hat{B}, \hat{D}\} $, and thus we can form an initial estimate $ \hat{\bm{\xi}} $. 
The first-order Taylor approximation of $ \mathbf{h}(\bm{\xi}) $ at the point $ \hat{\bm{\xi}} $ gives $ \mathbf{h}(\bm{\xi}) \approx \mathbf{h}(\hat{\bm{\xi}}) + (\mathbf{J} |_{\bm{\xi}=\hat{\bm{\xi}}}) \cdot (\bm{\xi}-\hat{\bm{\xi}}) $. Using this result, \eqref{eq:model_loc} can be rewritten as another linear equation set
\begin{equation}
	\mathbf{r} = \hat{\mathbf{J}} \Delta\bm{\xi} + \bm{\varsigma}, 
\end{equation}
where we have let $ \mathbf{r} = \hat{\bm{\eta}} - \mathbf{h}(\hat{\bm{\xi}}) $, $ \hat{\mathbf{J}} = \mathbf{J} |_{\bm{\xi}=\hat{\bm{\xi}}} $, and $ \Delta\bm{\xi} = \bm{\xi}-\hat{\bm{\xi}} $. 
Then MLE of $ \Delta\bm{\xi} $ is again equivalent to the following WLS estimation
\begin{equation}\label{eq:delta_xi}
	\Delta\hat{\bm{\xi}} = (\hat{\mathbf{J}}^{\mathsf{T}} \mathbf{W} \hat{\mathbf{J}})^{-1} \hat{\mathbf{J}}^{\mathsf{T}} \mathbf{W} \mathbf{r},
\end{equation}
where the weight matrix is $ \mathbf{W}(\bm{\eta}) = \bm{\Sigma}^{-1} (\bm{\eta}) $ and can be similarly approximated by $ \mathbf{W}(\hat{\bm{\eta}}) $. 
Once $ \Delta\hat{\bm{\xi}} $ is obtained, we renew the estimate from 
\begin{equation}\label{eq:xi_hat}
	\hat{\bm{\xi}} \leftarrow \hat{\bm{\xi}} + \Delta\hat{\bm{\xi}}.
\end{equation}
Then $ \Delta\hat{\bm{\xi}} $ can be re-obtained using \eqref{eq:delta_xi} with new $ \{ \mathbf{r},\hat{\mathbf{J}}\} $, so as to update $ \hat{\bm{\xi}} $.  
Repeat the procedure above to sequentially compensate for the estimation error. The convergence of our iterative algorithm requires only few iterations (i.e., normally two to four). 

After enough accurate iterations, $ \Delta \hat{\bm{\xi}} $ is given by
\begin{equation}
	\Delta \hat{\bm{\xi}} = (\mathbf{J}^{\mathsf{T}} \mathbf{W} \mathbf{J})^{-1} \mathbf{J}^{\mathsf{T}} \mathbf{W} \bm{\varsigma}.  
\end{equation}
The covariance matrix of the estimation error $ \Delta \hat{\bm{\xi}} $ above is $ (\mathbf{J}^{\mathsf{T}} \bm{\Omega}(\bm{\eta}) \mathbf{J})^{-1} $, equal to the inverse of FIM in \eqref{eq:FIM0}. 
Thus, the accuracy of our estimator can reach the CRLB under OMM. 

Our two-stage 3D-JPVE method using channel parameters is summarized in Algorithm~\ref{alg:loc}. 
Define $ N_1 = 4N $ and $ N_2 = 6N $ as the dimensions of measurements, and let $ E_1 = 6 $ and $ E_2 = 8 $ denote the dimensions of unknowns.  
The coarse estimator has a complexity of $ \mathcal{O}(I_4(N_1E_1^2+E_1^3)) $ with $I_4$ iterations, and the fine estimator costs a complexity of $ \mathcal{O}(I_5(N_2E_2^2+E_2^3)) $ with $I_5$ iterations. 
Since very few iterations (typically two to four for each stage) suffice, our estimation method has low complexity. 

\begin{algorithm}[t]
	\caption{Two-Stage 3D-JPVE Method Using Channel Parameter Estimates.}
	\renewcommand{\algorithmicrequire}{\textbf{Input:}}
	\renewcommand{\algorithmicensure}{\textbf{Output:}}
	\label{alg:loc}
	\begin{algorithmic}[1]
		\REQUIRE 
		Estimates  
		$ \{ \hat{d}_{n,1},\hat{d}_{n,2},\hat{r}_{n,1},\hat{r}_{n,2},\hat{\varphi}_{n,\mathrm{az}},\hat{\varphi}_{n,\mathrm{el}} \}_{n=1}^N $.
		\renewcommand{\algorithmicrequire}{\textbf{Implementation:}}
		\REQUIRE
		\STATE Form vectors $ \{ \hat{\mathbf{e}}_n, \hat{\mathbf{f}}_n, \hat{\mathbf{g}}_n \}_{n=1}^N $ and collections $ \{ \hat{\bm{\rho}}, \hat{\bm{\eta}} \} $. 
		\STATE Construct $ \hat{\bm{\Phi}} $ and let $ \mathbf{J}_{\mathrm{err}} \leftarrow \mathbf{I}_{4N} $. 
		\WHILE{{\it the stop criterion is not met}}
		\STATE Construct $ \mathbf{W}_{\mathrm{d}}(\hat{\bm{\eta}}, \hat{\bm{\theta}}) = ( \mathbf{J}_{\mathrm{err}}(\hat{\bm{\theta}}) \bm{\Sigma}_{\mathrm{d}}(\hat{\bm{\eta}}) \mathbf{J}_{\mathrm{err}}^{\mathsf{T}}(\hat{\bm{\theta}}) )^{-1}  $. 
		\STATE Compute the estimate $ \hat{\bm{\theta}} $ using \eqref{eq:theta_hat}.
		\ENDWHILE \;\,{\bf (Completion of Stage~I)}
		\STATE Compute $ \{\hat{B}, \hat{D}\} $ using $ \hat{\bm{\theta}} $ and form initial $ \hat{\bm{\xi}} $. 
		\STATE Construct $ \mathbf{W}(\hat{\bm{\eta}}) = \bm{\Sigma}^{-1} (\hat{\bm{\eta}}) $. 
		\WHILE{{\it the stop criterion is not met}}
		\STATE Compute $ \mathbf{r} $ and $ \hat{\mathbf{J}} $ using $ \hat{\bm{\xi}} $. 
		\STATE Compute $ \Delta \hat{\bm{\xi}} $ using \eqref{eq:delta_xi} and update $ \hat{\bm{\xi}} $ using \eqref{eq:xi_hat}. 
		\ENDWHILE \;\,{\bf (Completion of Stage~II)}
		
		\ENSURE 
		Estimates  
		$ \{ \hat{\mathbf{p}}, \hat{\mathbf{v}}, \hat{B}, \hat{D} \} $.
	\end{algorithmic}
\end{algorithm}

\section{Numerical Results}\label{sec:NR}

\subsection{Simulation Configurations}

Our simulation experiments use the following settings unless otherwise specified. For the setup, let $N = 3$ snapshots with $t_{n1} = 0.2(n-1), \, n=1,\ldots,N$ in seconds.  
The system bandwidth is $B = 240\,\text{MHz}$ with a central carrier frequency of $28\,\text{GHz}$.  
Each frame transmits $G = 64$ OFDM symbols containing pilots, where $G_1 = G_2 = \sqrt{G}$.  
The total number of subcarriers is $K_0 = 2048$, of which only the first $K = 200$ are allocated for pilots.  
The transmit power of BS is set to $ \mathbf{P}_{\mathrm{T}} = $~20\,dBm, and the power of active RIS is also $ \mathbf{P}_{\mathrm{R}} =$~20\,dBm with the amplification coefficient $ \eta $ computed from \cite[Eq.~(1)]{Zheng2024JrCUP}. 
The array size of RIS is $ M_{\mathrm{x}} = M_{\mathrm{y}} = 15 $ and the element spacing is $ \Delta_{\mathrm{s}} = 0.2\lambda $. 
Concerning the complex path gain, the amplitude is generated by $ \lambda/4\pi d $ with the transmission distance $ d $, and the phase is randomly drawn from $ \mathcal{U}(0,2\pi) $. 

The BS is located at $ \mathbf{q}_1 = [30, 30, 0]^{\mathsf{T}} $\,m, and the RIS is positioned at $ \mathbf{q}_1 = [0, 0, 0]^{\mathsf{T}} $\,m with rotation matrix $ \mathbf{R} = \mathbf{I}_3 $. 
The UE's default initial position and velocity are $ \mathbf{p} = [-25, 42, -15]^{\mathsf{T}} $\,m and $ \mathbf{v} = [-25, 25, 0]^{\mathsf{T}} $\,m/s, respectively, and we will also study the case with varying $ \{\mathbf{p},\mathbf{v}\} $. 
We set the initial clock bias and clock drift to $ B = $~100\,ns and $ D = $~5\,parts per million (ppm), respectively. 

The received signal-to-noise ratio (SNR) over $ N $ snapshots is defined as 
\begin{equation}
	\operatorname{SNR} = 
	\frac{ \sum_{n=1}^N \| \mathbf{Y}_{\mathrm{L}}^{(n)} + \mathbf{Y}_{\mathrm{R}}^{(n)}\|^2_{\mathsf{F}} }
	{ \sum_{n=1}^N \| \mathbf{N}^{(n)}\|^2_{\mathsf{F}}  }. 
\end{equation} 
The default SNR is set to 15\,dB. 

\subsection{Evaluation Settings}

The CRLB derived in Section~\ref{sec:CRLB} indicates the achievable optimal estimation accuracy, and the performance of benchmark schemes is also provided for comparison. 

Concerning the CE step, our method competes with (i) the multiple iterations and successive cancellation (MI+SC) method \cite{Keykhosravi2022RISenabled}, also consisting of coarse and fine estimation stages, and (ii) the VSCPD and the following mode-wise optimization (VSCPD+MWO) method \cite{Zhang2022Tensor,Zhang2024Integrated}. 
Notably, the MI+SC method employs different RIS profile from ours and hence the CRLB for channel parameters differs. 
The algorithm parameters of MI+SC, e.g., the number of DFT points and candidate AODs, are maintained at the same values as those in \cite{Keykhosravi2022RISenabled}.  
In addition, the VSCPD+MWO method was originally tailored for the 2D-JADE issue, with only a single AOD to be estimated; we extend it to our 3D-JADE case by using the proposed tensor formulation scheme. 
Herein, parameters are individually extracted in each mode using the VSCPD results, and the pseudorange rates are finally determined from the average result of those in modes two and three with random initialization. 

Further, as the second step in 3D-JPVE is a specialized problem, there is no existing solution for comparison. 
Consequently, we design two ideal schemes under our framework to show the optimal performance of our two stages: (i) the ground truth $ \bm{\theta} $ is used to construct the weight matrix $ \mathbf{W}_{\mathrm{d}} $ for the coarse estimation, and (ii) the ground truth $ \bm{\xi} $ is initially fed to our iterative algorithm for the fine estimation. 
Moreover, we also simulate the conventional, degraded single-epoch localization (SEL) case, where only the single-epoch (i.e., the first-epoch) measurements  excluding Doppler shifts are leveraged to estimate 3D position and clock bias (see {\it Scheme~A} in Section~\ref{sec:feasible}).  
Herein, our degraded method with merely Stage~I required (to be shown at Section~\ref{sec:loc_N}) competes with the search-based localization (SBL) method \cite{Keykhosravi2022RISenabled}. 
In SBL, the RIS-UE distance is exhaustively searched to obtain the global optimal solution of a nonconvex optimization problem.

We use the root mean squared error (RMSE) to evaluate the estimation accuracy of scalar and vector parameters, defined as
\begin{equation}
\operatorname{RMSE}(\bm{\chi}) = 
\sqrt{ \mathbb{E}  \big[ \| \hat{\bm{\chi}} - \bm{\chi} \|^2 \big] }, 
\end{equation}
where $ \bm{\chi} $ can be any channel and state parameters, and $ \hat{\bm{\chi}} $ denotes the estimate. 
Further, to assess the comprehensive performance of channel parameter estimation, we define the normalized MSE (NMSE) for channel matrix estimation as
\begin{equation}
	\operatorname{NMSE}( \mathbf{H} ) = 
	\frac{ \mathbb{E} \big[  \big\| \mathbf{H} ( \hat{\bm{\eta}}_{\mathrm{ch}} ) -
		\mathbf{H} ( \bm{\eta}_{\mathrm{ch}} ) \|^2_{\mathsf{F}}  \big] }
	{ \| \mathbf{H} ( \bm{\eta}_{\mathrm{ch}} ) \|^2_{\mathsf{F}} } ,  
\end{equation}
where $ \mathbf{H} ( \bm{\eta}_{\mathrm{ch}} ) \in \mathbb{C}^{K \times G} $ denotes the channel matrix at a single snapshot, found from the received signal matrix divided by the transmit pilot. 
To accurately compute the statistical expectation, we perform at least $ 10^3 $ Monte-Carlo trials for each snapshot. 
Both RMSE and NMSE during CE step are obtained from the average result over $ N $ snapshots.  

\begin{figure*}[t]
	\centering
	\includegraphics[width=0.97\linewidth]{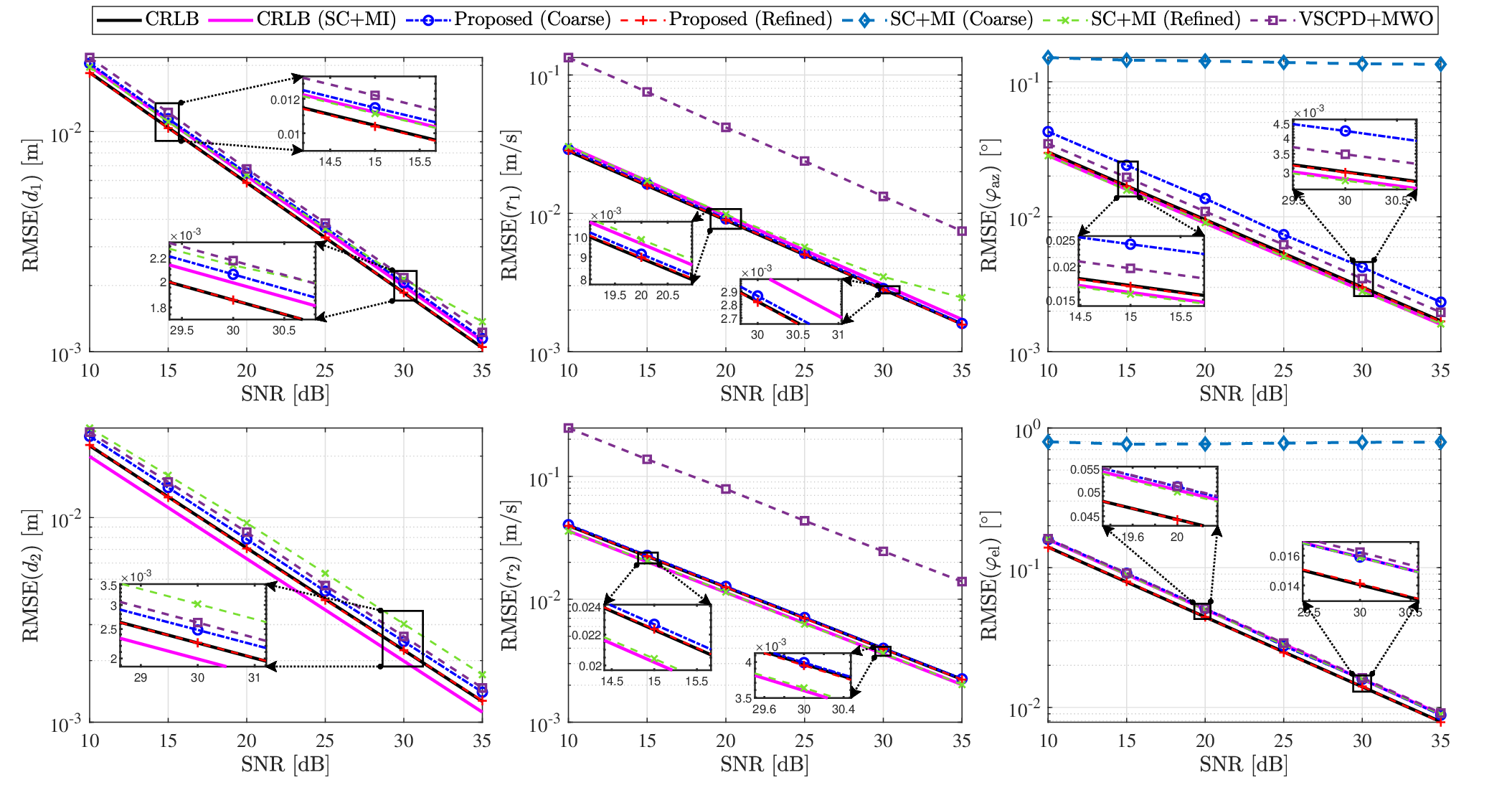}
	\caption{RMSE of channel parameters $ d_1 $, $ d_2 $, $ r_{1} $, $ r_{2} $, $ \varphi_{\mathrm{az}} $, and $ \varphi_{\mathrm{el}} $ versus SNR.} 
\label{fig:RMSE_CE}
\end{figure*}

\subsection{CE Performance}

\subsubsection{Estimation of Channel Parameters}\label{sec:CE_RMSE}
The RMSE of channel parameter estimation versus SNR is presented in Fig.~\ref{fig:RMSE_CE}. The DFT-based coarse SC+MI algorithm exhibits poor overall accuracy, so only its angle estimation results are shown for visual clarity. The refined SC+MI approach attains CRLB-level estimates for $\{ r_2, \varphi_{\mathrm{az}}, \varphi_{\mathrm{el}} \}$, but remains suboptimal for $\{ d_1, d_2, r_1 \}$ especially at high SNR, due to strong parameter coupling. By contrast, the proposed coarse CE method significantly outperforms the coarse SC+MI baseline. Our subsequent MLE-based refinement further improves accuracy, consistently achieving the CRLB for all parameters. In addition, the VSCPD+MWO method yields inaccurate estimates of $\{ r_1, r_2 \}$ because the scaling coefficient of the unknown in mode~3 is much smaller than in mode~2, reducing the precision of the mode-3 estimate and hence the averaged result. Finally, although our coarse CE approach and VSCPD+MWO follow similar procedures for the estimation of $\{ d_1, d_2, \varphi_{\mathrm{az}}, \varphi_{\mathrm{el}} \}$, their performance differs. This can be attributed to the ALS refinement of VSCPD results: ALS improves factor estimates in most modes at the expense of reduced accuracy in very few modes.

\begin{figure}[t]
	\centering
	\includegraphics[width=0.91\linewidth]{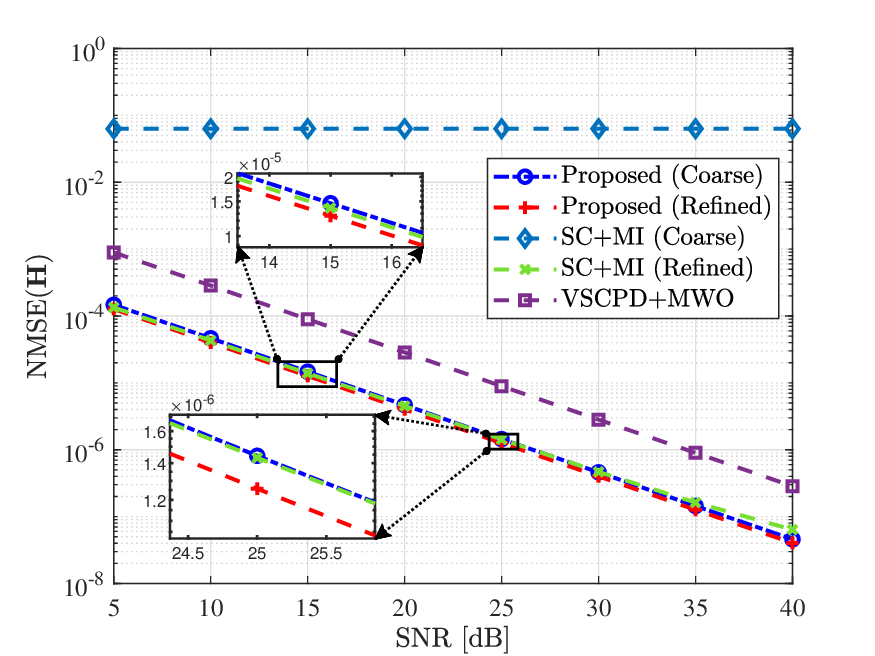}
	\caption{NMSE of channel matrix $\mathbf{H}$ versus SNR.} 
	\label{fig:NMSE_SNR}
\end{figure}

\subsubsection{Estimation of Channel Matrix}

The NMSE of channel matrix versus SNR is illustrated in Fig.~\ref{fig:NMSE_SNR} as the comprehensive performance metric of channel parameter estimation. 
The performance of coarse SC+MI estimation method is inferior, surpassed by the VSCPD+MWO method with significant improvement. 
Our coarse CE method shares similar performance with the SC+MI algorithm of the refined estimation stage, further outperforming VSCPD+MWO.  
Finally, the proposed MLE-based fine method achieves the best estimation precision.

\subsection{3D-JPVE Performance}\label{sec:sim_loc}

\begin{figure}[t]
	\centering
	\includegraphics[width=0.91\linewidth]{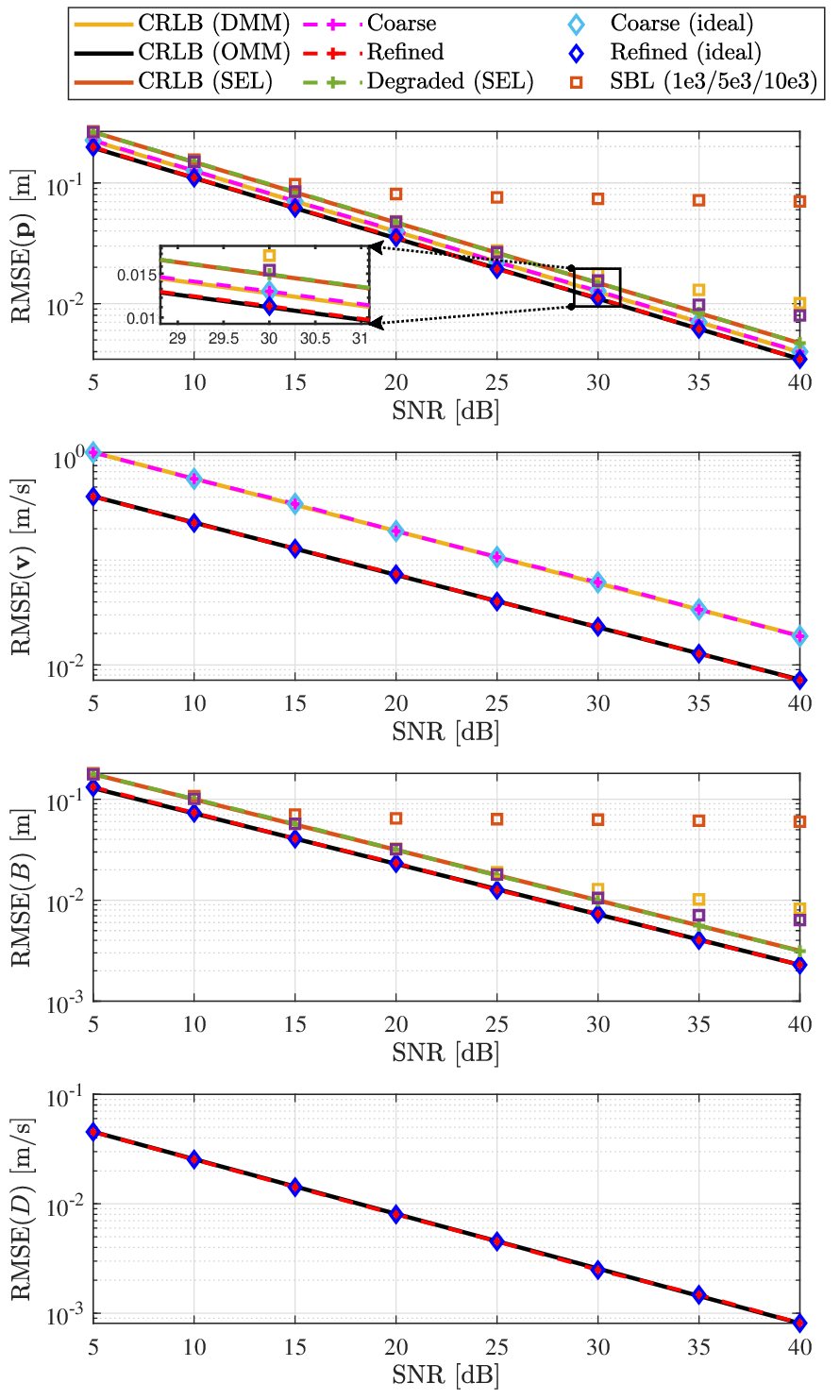}
	\caption{RMSE of state parameters $\mathbf{p}$, $\mathbf{v}$, $B$, and $D$ versus SNR. In SBL, 
	the search number within the distance range $(0,150)$\,m is set to $10^3$ (orange square), $ 5\times10^3 $ (yellow), and $ 10^4 $ (purple).  
	} 
	\label{fig:loc_SNR}
\end{figure}

\subsubsection{Varying SNR}
Figure~\ref{fig:loc_SNR} illustrates the localization and synchronization performance of a high-mobility UE under varying SNR. The SBL scheme is evaluated by searching over the distance range $(0,150)$\,m with search numbers $10^3$ (orange square), $5\times10^3$ (yellow), and $10^4$ (purple). The CRLB under OMM is consistently lower than that under DMM, confirming the information loss introduced by multi-epoch differential measurements. Both the coarse and fine estimators achieve their corresponding CRLBs, demonstrating the statistical efficiency of the proposed 3D-JPVE method. Even in the degraded SEL case, our method attains CRLB-level accuracy. In contrast, the SBL algorithm requires significantly large search numbers to approach high accuracy, incurring heavy computational overhead, whereas the proposed low-complexity approach reaches optimal accuracy with only very few iterations (typically two).

\begin{figure}[t]
	\centering
	\includegraphics[width=0.9\linewidth]{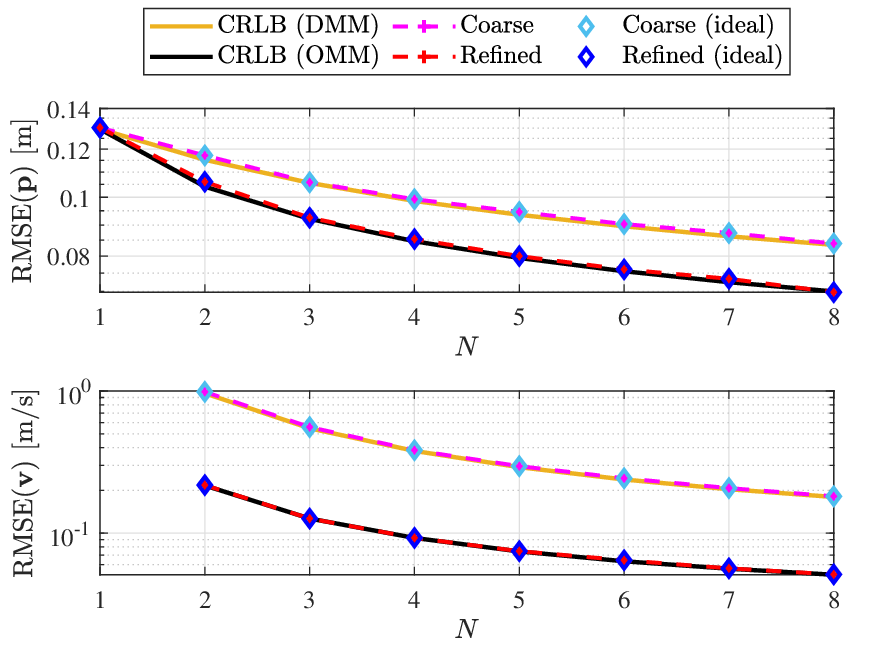}
	\caption{RMSE of position $\mathbf{p}$ and velocity $\mathbf{v}$ versus snapshot number $N$.} 
	\label{fig:loc_Num}
\end{figure}

\subsubsection{Varying $ N $}\label{sec:loc_N}
Figure~\ref{fig:loc_Num} shows the position and velocity estimation performance under different snapshot number $N$'s. While velocity estimation requires at least two snapshots, the position determination is always feasible regardless of $N$.   As $N$ increases, the estimation accuracy of both $ \mathbf{p} $ and $ \mathbf{v} $ improves, implying that more reliable position and velocity estimates are achievable at the expense of less time efficiency.  The OMM-based fine estimator outperforms the DMM-based coarse estimator, except for the SEL case with $N=1$, where their position estimates coincide. Notably, both stages of our estimator attain CRLB-level performance, confirming the statistical efficiency of the proposed method in varying-$N$ scenarios.

\subsubsection{Varying $ \{K,G\} $ With Different $ \{\mathbf{p},\mathbf{v}\} $ }

\begin{figure}[t]
	\centering
	\includegraphics[width=0.92\linewidth]{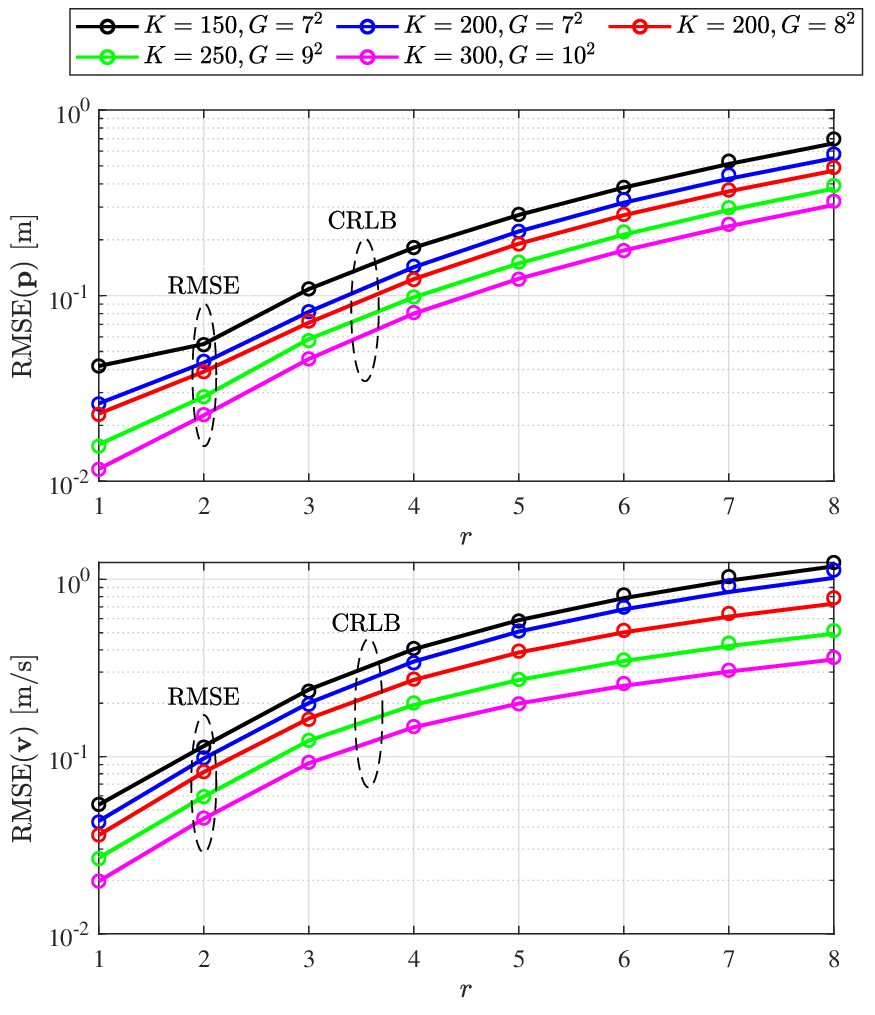}
	\caption{RMSE of position $\mathbf{p}$ and velocity $\mathbf{v}$ versus parameter $r$ with different number of subcarriers ($K$) and OFDM symbols ($G$) for pilot transmssion. The position and velocity are set to $ \mathbf{p} = [-15r/\sqrt{2}, 15r/\sqrt{2}, -10]^{\mathsf{T}} $\,m and $ \mathbf{v} = [-5r/\sqrt{2}, 5r/\sqrt{2}, 0]^{\mathsf{T}} $\,m/s, respectively, with $r$ varying from 1 to 8. The solid line and the discrete circle represent the OMM-based CRLB and RMSE, respectively.  } 
	\label{fig:loc_pv+KG}
\end{figure}

The estimation accuracy with varying numbers $ \{K,G\} $ and states $ \{\mathbf{p},\mathbf{v}\} $ is plotted in Fig.~\ref{fig:loc_pv+KG}.  
We design the UE's initial position and velocity as $ \mathbf{p} = [-15r/\sqrt{2}, 15r/\sqrt{2}, -10]^{\mathsf{T}} $\,m and $ \mathbf{v} = [-5r/\sqrt{2}, 5r/\sqrt{2}, 0]^{\mathsf{T}} $\,m/s, respectively, for $r=1,\ldots,8$. 
The solid line represents the CRLB under OMM, and the discrete circle represents the RMSE from our fine estimator. 
We observe that the estimation error grows with increasing $r$, primarily due to the inherent degradation of angle-based position and velocity estimation over wide-range scenarios. Fortunately, increasing $K$ and/or $G$ improves accuracy, helping to counteract this deterioration above. Moreover, the proposed estimator achieves CRLB-level performance in most cases, with only slight deviations at larger $r$ values.

\subsubsection{Active RIS v.s. Passive RIS}\label{sec:RIS_comp}
\begin{figure}[t]
	\centering
	\includegraphics[width=0.92\linewidth]{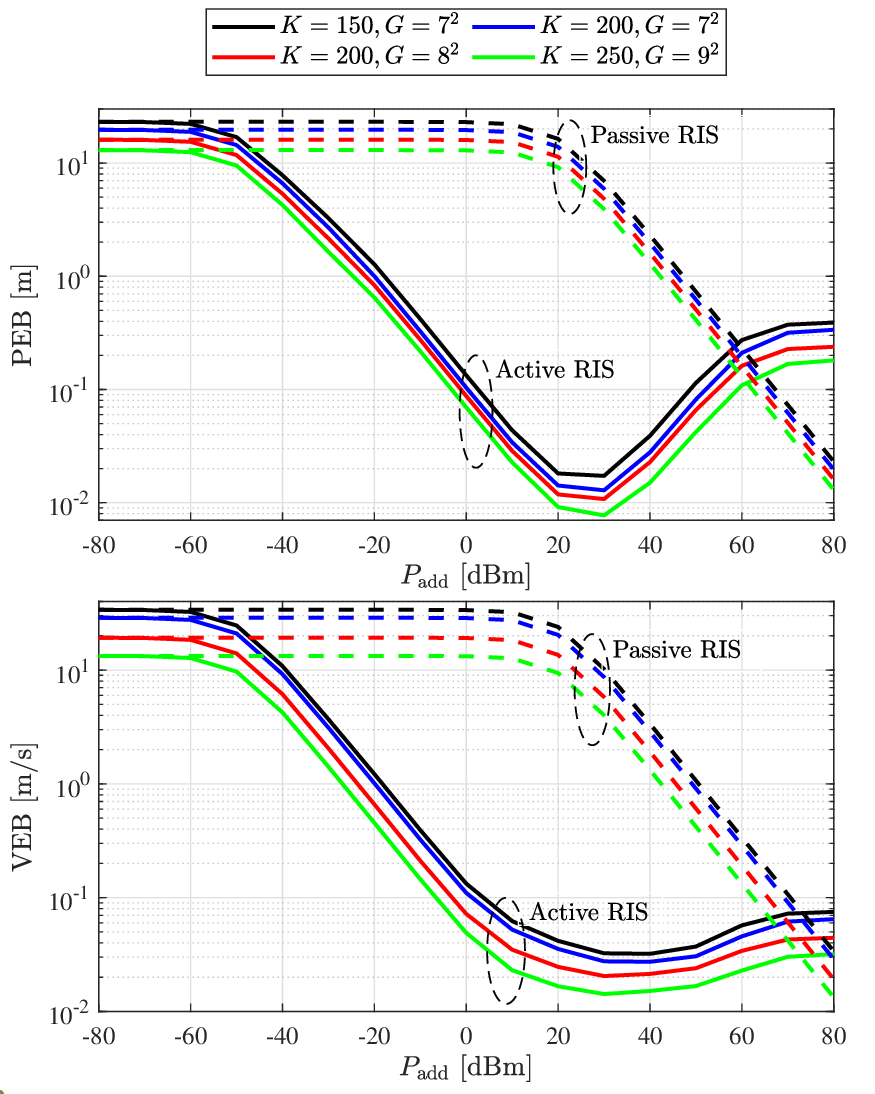}
	\caption{PEB and VEB versus additional system power $ P_{\mathrm{add}} $ in the active RIS case (solid line) and the passive RIS case (dashed line). Results are presented for different combinations of the number of subcarriers ($K$) and OFDM symbols ($G$), and each color of line represents one combination for both cases.  } 
	\label{fig:loc_pow}
\end{figure}

Recall that, unlike the passive RIS, the active RIS can amplify the incident signal using external power, but at the cost of introducing additional thermal noise. To assess their overall impact on estimation performance, Fig.~\ref{fig:loc_pow} compares the PEB and VEB for different RIS types under different power budgets and pilot numbers. The solid and dashed lines are from the active RIS and passive RIS cases, respectively, and each color of line represents one combination of $\{K, G\}$ for both cases. In the active RIS scenario, the BS transmit power is fixed at $P_{\mathrm{T}}=20$\,dBm, while the RIS power is set to $P_{\mathrm{R}}=P_{\mathrm{add}}$, where $P_{\mathrm{add}}$ represents the additional system power. For a fair comparison under the same total system power, the passive RIS case uses a transmit power of $P_{\mathrm{T}}+P_{\mathrm{add}}$, with $P_{\mathrm{R}}=0$ and $\sigma_{\mathrm{R}}=0$.

From Fig.~\ref{fig:loc_pow}, it is observed that in the passive RIS case, the estimation error decreases slightly with small increases in $P_{\mathrm{add}}$, but drops sharply once $P_{\mathrm{add}} > 10$\,dBm. This trend is expected since higher power budgets improve the received SNR at a similar rate. In contrast, the error bound for the active RIS first decreases and then increases, with optimal performance occurring around $P_{\mathrm{add}} \approx 30$\,dBm. The initial improvement stems from the growing power balance between the direct BS–UE and cascaded BS–RIS–UE links. In contrast, excessive RIS power eventually disrupts this balance, and also amplifies the received RIS-incurred noise. Although passive RIS may surpass active RIS at unrealistically high power levels (beyond 60\,dBm), active RIS consistently offers substantial gains within practical power ranges ($-30$\,dBm $< P_{\mathrm{add}} < 40$\,dBm). Moreover, increasing $K$ and/or $G$ enhances estimation accuracy for both RIS types.

\section{Conclusion}\label{sec:conc} 
This paper investigated optimal joint 3D position and velocity estimation (3D-JPVE) for high-mobility, unsynchronized UEs in SISO systems with a single BS. It was theoretically shown that the estimation problem becomes infeasible when relying solely on a single RIS or on multi-snapshot measurements alone. To address this, a realizable solution was proposed by jointly leveraging RIS deployment and multi-epoch measurements. A two-stage multi-snapshot channel estimation framework was developed, where a third-order tensor formulation enabled coarse parameter estimation, followed by maximum likelihood–based fine estimation. Building on the estimated channel parameters, a second two-stage scheme for 3D-JPVE was designed, combining differential measurement–based coarse estimation with iterative refinement to attain optimality. Furthermore, closed-form CRLB expressions were derived, and both estimation errors and computational complexity were theoretically analyzed. Simulation results validated the effectiveness of the proposed estimators and highlighted the performance advantages of active RIS over passive RIS. Overall, the study established a viable framework for optimal estimation of both channel parameters and UE states, while offering new insights into the enabling role of RIS deployment and multi-snapshot utilization for 3D-JPVE  in high-mobility scenarios.


\appendices

\renewcommand\thesubsectiondis{\thesection.\arabic {subsection}}

\section{Proof of Infeasibility of {\it Scheme B}}\label{sec:proof}

\subsection{Overall Proof Procedure}

Channel estimation yields
\begin{equation}\label{eq:rho_woR}
	\hat{\bm{\eta}}_{\mathrm{woR}} = \mathbf{h}_{\mathrm{woR}} ( \bm{\xi} ) + \bm{\varsigma}_{\mathrm{woR}}, 
\end{equation}
where the mapping function vector $ \mathbf{h}_{\mathrm{woR}}( \bm{\xi} ) $ is given by
\begin{equation}
	\mathbf{h}_{\mathrm{woR}} ( \bm{\xi} ) = [ \ldots,d_{n,1},r_{n,1} ,\ldots ]^{\mathsf{T}} \in\mathbb{R}^{2N\times 1},
\end{equation}
vectors $ \hat{\bm{\eta}}_{\mathrm{woR}} $ and $ \bm{\varsigma}_{\mathrm{woR}} $ are the estimated version of $ \mathbf{h}_{\mathrm{woR}} $ and the estimation error, respectively. 
The Taylor expansion of $ \mathbf{h}_{\mathrm{woR}} ( \bm{\xi} ) $ at an estimate point $ \hat{\bm{\xi}} $ is expressed as
\begin{equation}\label{eq:h_woR}
	\mathbf{h}_{\mathrm{woR}} ( \bm{\xi} ) = \mathbf{h}_{\mathrm{woR}} ( \hat{\bm{\xi}} ) + (\mathbf{J}_{\mathrm{woR}}|_{\bm{\xi}=\hat{\bm{\xi}}} ) \cdot ( \bm{\xi} - \hat{\bm{\xi}} ) + o(\|\Delta \bm{\xi}\|),
\end{equation}
where $ \mathbf{J}_{\mathrm{woR}} =  \partial \mathbf{h}_{\mathrm{woR}} ( \bm{\xi} ) / \partial \bm{\xi}^{\mathsf{T}}  \in\mathbb{R}^{2N\times 8} $ denotes the Jacobian matrix, whose elements are presented Appendix~\ref{sec:Derivatives}.2. 
Substituting \eqref{eq:h_woR} into \eqref{eq:rho_woR} comes to
\begin{equation}\label{eq:r_woR}
	\mathbf{r}_{\mathrm{woR}} = \hat{\mathbf{J}}_{\mathrm{woR}} \Delta\bm{\xi} + \bm{\varsigma}_{\mathrm{woR}},
\end{equation}
where $ \mathbf{r}_{\mathrm{woR}} = \hat{\bm{\eta}}_{\mathrm{woR}} - \mathbf{h}_{\mathrm{woR}} ( \hat{\bm{\xi}} ) - o(\|\Delta \bm{\xi}\|) $,  
$ \hat{\mathbf{J}}_{\mathrm{woR}} = \mathbf{J}_{\mathrm{woR}}|_{\bm{\xi}=\hat{\bm{\xi}}} $, and $ \Delta \bm{\xi} = \bm{\xi} - \hat{\bm{\xi}} $. 
Consequently, the estimation of $ \bm{\xi} $ using \eqref{eq:rho_woR} is equivalent to the acquisition of $ \Delta\bm{\xi} $ using \eqref{eq:r_woR}. 
Nevertheless, per Appendix~\ref{sec:proof}.2, $ \hat{\mathbf{J}}_{\mathrm{woR}} $ is a rank-deficient matrix when $ N\ge4 $ (the rank-deficiency even also exists in the clock-synchronized case). 
Therefore, the estimation of $ \Delta\bm{\xi} $ using \eqref{eq:r_woR} is an underdetermined problem, reflecting the infeasibility of {\it Scheme B} for 3D-JPVE. 

\subsection{Proof of the Rank-Deficiency of $ \hat{\mathbf{J}}_{\mathrm{woR}} $}

We first point out that the proof of the rank-deficiency of the estimate $ \hat{\mathbf{J}}_{\mathrm{woR}} $ is same to that of the ground truth $ \mathbf{J}_{\mathrm{woR}} $, and transposing a matrix does not alter the rank. 
Therefore, for notational simplicity, we demonstrate as follows that $ \operatorname{rank} (\mathbf{G}_{\mathrm{woR}}) < 8 $ when $ N\ge4 $, where we have let $ \mathbf{G}_{\mathrm{woR}} = \mathbf{J}_{\mathrm{woR}}^{\mathsf{T}} \in\mathbb{R}^{8\times 2N} $. 

Define two vectors $ \mathbf{g}_{n,\mathrm{od}} =  [\mathbf{G}_{\mathrm{woR}}]_{:,2\ell-1} $ and $ \mathbf{g}_{n,\mathrm{ev}} =  [\mathbf{G}_{\mathrm{woR}}]_{:,2\ell},\,n,\ell=1,\ldots,N $ as the odd- and even-indexed columns of $ \mathbf{G}_{\mathrm{woR}} $, respectively. 
Vectors $ \mathbf{g}_{n,\mathrm{od}} $ and $ \mathbf{g}_{n,\mathrm{ev}} $ can be expressed using Appendix~\ref{sec:Derivatives}.2 as 
\begin{align}\label{eq:gn2}
	\mathbf{g}_{n,\mathrm{od}} & = 
	\big[ -\mathbf{k}_{n,1},\, -t_{n1} \mathbf{k}_{n,1},\, 1,\,  t_{n1} \, \big]^{\mathsf{T}} ,
	\nonumber \\
	\mathbf{g}_{n,\mathrm{ev}} &= \big[\, \dot{\mathbf{k}}_{n,1},\, \mathbf{k}_{n,1} + t_{n1} \dot{\mathbf{k}}_{n,1},\, 0,\, 1 \,\big]^{\mathsf{T}} ,
\end{align}
where we have let $ \dot{\mathbf{k}}_{n,1} = \partial \mathbf{k}_{n,1} / \partial t_{n1} $. 

Further define $ \tilde{\mathbf{g}}_{n,\mathrm{od}} = [\mathbf{g}_{n,\mathrm{od}}]_{1:6} $ and $ \tilde{\mathbf{g}}_{n,\mathrm{ev}} = [\mathbf{g}_{n,\mathrm{ev}}]_{1:6} $. 
Note that  
\begin{align}
	\tilde{\mathbf{g}}_{n,\mathrm{od}} &= 
	-s_{n,1} \big[ \mathbf{u}, t_{n1} \mathbf{u} \big]^{\mathsf{T}} + 
	t_{n1}s_{n,1} \big[ \mathbf{v}, t_{n1} \mathbf{v} \big]^{\mathsf{T}},
	\nonumber \\
	\tilde{\mathbf{g}}_{n,\mathrm{ev}} &= 
	z_{n,1} \big[ \mathbf{u}, t_{n1} \mathbf{u} \big]^{\mathsf{T}} + 
	c_{n,1} \big[ \mathbf{v}, t_{n1} \mathbf{v} \big]^{\mathsf{T}}
	\nonumber \\
	&+s_{n,1} \big[ \mathbf{0}_{3\times 1}, \mathbf{u} \big]^{\mathsf{T}} - 
	t_{n1} s_{n,1} \big[\mathbf{0}_{3\times 1}, \mathbf{v} \big]^{\mathsf{T}},
\end{align}
where $ \mathbf{u} = \mathbf{q}_1 - \mathbf{p} $, $ z_{n,1} = s^3_{n,1} \mathbf{b}_{n,1}^{\mathsf{T}}\mathbf{v} $, and $ c_{n,1} = -t_{n1}z_{n,1}- s_{n,1} $. 
Therefore, the column space  $ \mathcal{S} = \operatorname{Col} ([\mathbf{G}]_{1:6,:})  $ when $ N\ge4 $ is given by
\begin{align}
	\mathcal{S} &= \operatorname{Span} \big\{ 
		\big[ \mathbf{u}, t_{n1} \mathbf{u} \big]^{\mathsf{T}},
		\big[ \mathbf{v}, t_{n1} \mathbf{v} \big]^{\mathsf{T}},
		\big[ \mathbf{0}_{3\times 1}, \mathbf{u} \big]^{\mathsf{T}},
		\big[\mathbf{0}_{3\times 1}, \mathbf{v} \big]^{\mathsf{T}}
	  \big\} 
	\nonumber \\
	&= \operatorname{Span} \big\{ 
		\big[ \mathbf{u}, \mathbf{0}_{3\times 1} \big]^{\mathsf{T}},
		\big[ \mathbf{v}, \mathbf{0}_{3\times 1} \big]^{\mathsf{T}},
		\big[ \mathbf{0}_{3\times 1}, \mathbf{u} \big]^{\mathsf{T}},
		\big[\mathbf{0}_{3\times 1}, \mathbf{v} \big]^{\mathsf{T}}
	  \big\}. 
\end{align}
In general, vectors $\mathbf{u}$ and $\mathbf{v}$ are linearly independent (if not, the rank becomes even smaller). Thus, we obtain  $ \operatorname{dim}(\mathcal{S}) = 4 $, or equivalently, $ \operatorname{rank}([\mathbf{G}]_{1:6,:}) = 4 $, when $ N\ge4 $.
This indicates that the tall Jacobian matrix is rank-deficient even when the clock is synchronized (i.e., $ B=0 $ and $ D=0 $). 

Further, notice that $ \operatorname{rank}( [\mathbf{A},\mathbf{B}] ) \le \operatorname{rank}( \mathbf{A} ) + \operatorname{rank}( \mathbf{B} ) $ for arbitrary matrices $ \{ \mathbf{A},\mathbf{B} \} $. 
With \eqref{eq:gn2}, it is straightforward to find that $ \operatorname{rank} ( [\mathbf{G}_{\mathrm{woR}}]_{7:8,:} ) = 2 $. 
Combining the rank results regarding the upper and lower submatrices, we finally establish $  \operatorname{rank} (\mathbf{G}_{\mathrm{woR}}) \le 6 $  when $ N\ge4 $, showing the rank-deficiency of the tall Jacobian matrix $ \hat{\mathbf{J}}_{\mathrm{woR}} $.

\section{Partial Derivatives in Jacobian matrices}\label{sec:Derivatives}

\subsection{Derivatives in CE Domain}\label{sec:Derivatives_CE}

We first define 
\begin{gather}
    y_{\mathrm{L}}^{(n,g,k)} = [\mathbf{Y}_{\mathrm{L}}^{(n)}]_{k,g},\;
    y_{\mathrm{R}}^{(n,g,k)} = [\mathbf{Y}_{\mathrm{R}}^{(n)}]_{k,g},
    \nonumber \\
	d_{0} = \| \mathbf{q}_1 - \mathbf{q}_2 \|,\;
	\bm{\varphi} = [ \varphi_{\mathrm{az}}, \varphi_{\mathrm{el}} ]^{\mathsf{T}}, 
	\nonumber \\
	\varOmega^{(k)} = -\frac{2\pi}{c}F^{(k)},\; \varTheta^{(g)} = \frac{2\pi}{\lambda} T^{(g)},
	\nonumber \\
	\delta_{1}^{(n,k)} = e^{\jmath \varOmega^{(k)} d_{n,1} }, \;
	\delta_{2}^{(n,k)} = e^{\jmath \varOmega^{(k)} ( d_{n,2} + d_{0} )}, \; 
	\nonumber \\
	\vartheta_{1}^{(n,g)} = e^{\jmath \varTheta^{(g)} r_{n,1}}, \;
	\vartheta_{2}^{(n,g)} = e^{\jmath \varTheta^{(g)} r_{n,2}},
	\nonumber \\
	\dot{\delta}_{i}^{(n,k)} = \jmath \varOmega^{(k)} \delta_{i}^{(n,k)}, \;
	\dot{\vartheta}_{i}^{(n,g)} = \jmath \varTheta^{(g)} \vartheta_{i}^{(n,g)},\; i = 1,2,
	\nonumber \\
	\rho_{\mathrm{R}}^{(n,g)} = (\bm{\gamma}^{(n,g)})^{\mathsf{T}} \breve{\mathbf{a}}_{\mathrm{R}}(\bm{\psi}_{\mathrm{R}}^{(n)}),
	\nonumber \\
	\dot{\mathbf{e}}_{\mathrm{az}}(\bm{\varphi}) = 
	\big[ -\sin(\varphi_{\mathrm{az}}) \cos(\varphi_{\mathrm{el}}), \cos(\varphi_{\mathrm{az}}) \cos(\varphi_{\mathrm{el}}), 0 \big]^{\mathsf{T}}, 
	\nonumber \\
	\dot{\mathbf{e}}_{\mathrm{el}}(\bm{\varphi}) \! = \!
	\big[\! -\!\cos(\varphi_{\mathrm{az}}) \sin(\varphi_{\mathrm{el}}), \!-\!\sin(\varphi_{\mathrm{az}}) \sin(\varphi_{\mathrm{el}}), \cos(\varphi_{\mathrm{el}}) \big]^{\mathsf{T}}, 
	\nonumber \\
	\!\!\!
	\dot{\rho}_{\mathrm{R},\mathrm{az(el)}}^{(n,g)} = \jmath \frac{2\pi}{\lambda}
	\big(\bm{\gamma}^{(n,g)}\big)^{\mathsf{T}} \big( \breve{\mathbf{a}}_{\mathrm{R}}(\bm{\psi}_{\mathrm{R}}^{(n)}) \circledast (\mathbf{P}_{\mathrm{R}} \dot{\mathbf{e}}_{\mathrm{az(el)}}(\bm{\varphi}_n)) \big). \!\!\!
	\label{eq:deriv_defi}
\end{gather}
Then the derivatives in CE domain can be expressed as 
\begin{align}
	\frac{ \partial y_{\mathrm{L}}^{(n,g,k)} }{ \partial d_{n,1} } &= 
	\beta_{\mathrm{L}}^{(n)} 
	\dot{\delta}_{1}^{(n,k)}  \vartheta_{1}^{(n,g)},\; 
	\frac{ \partial y_{\mathrm{L}}^{(n,g,k)} }{ \partial r_{n,1} } = 
	\beta_{\mathrm{L}}^{(n)} 
	\delta_{1}^{(n,k)}  \dot{\vartheta}_{1}^{(n,g)},
	\nonumber \\
	\frac{ \partial y_{\mathrm{R}}^{(n,g,k)} }{ \partial d_{n,2} } &= 
	\beta_{\mathrm{R}}^{(n)} 
	\dot{\delta}_{2}^{(n,k)}  \vartheta_{2}^{(n,g)} \rho_{\mathrm{R}}^{(n,g)},
	\nonumber \\
	\frac{ \partial y_{\mathrm{R}}^{(n,g,k)} }{ \partial r_{n,2} } &= 
	\beta_{\mathrm{R}}^{(n)} 
	\delta_{2}^{(n,k)}  \dot{\vartheta}_{2}^{(n,g)} \rho_{\mathrm{R}}^{(n,g)},
	\nonumber \\
	\frac{ \partial y_{\mathrm{R}}^{(n,g,k)} }{ \partial \varphi_{n,\mathrm{az(el)}} } &= 
	\beta_{\mathrm{R}}^{(n)} 
	\delta_{2}^{(n,k)}  \vartheta_{2}^{(n,g)} \dot{\rho}_{\mathrm{R},\mathrm{az(el)}}^{(n,g)},
	\nonumber \\
	\frac{ \partial y_{\mathrm{L}}^{(n,g,k)} }{ \partial \alpha_{\mathrm{L},\mathfrak{R}}^{(n)} } &= 
	x \delta_{1}^{(n,k)}  \vartheta_{1}^{(n,g)}, \;
	\frac{ \partial y_{\mathrm{L}}^{(n,g,k)} }{ \partial \alpha_{\mathrm{L},\mathfrak{I}}^{(n)} } = \jmath
	\frac{ \partial y_{\mathrm{L}}^{(n,g,k)} }{ \partial \alpha_{\mathrm{L},\mathfrak{R}}^{(n)} },
	\nonumber \\
	\frac{ \partial y_{\mathrm{R}}^{(n,g,k)} }{ \partial \alpha_{\mathrm{R},\mathfrak{R}}^{(n)} } &= 
	x \alpha_{\mathrm{R},1} \delta_{2}^{(n,k)}  \vartheta_{2}^{(n,g)}  \!\rho_{\mathrm{R}}^{(n,g)},
	\frac{ \partial y_{\mathrm{R}}^{(n,g,k)} }{ \partial \alpha_{\mathrm{R},\mathfrak{I}}^{(n)} } = \jmath
	\frac{ \partial y_{\mathrm{R}}^{(n,g,k)} }{ \partial \alpha_{\mathrm{R},\mathfrak{R}}^{(n)} }.
	\nonumber \\
\end{align}
Note that for complex-valued function $ \mathbf{B}(\phi) $ with respect to real-valued scalar $ \phi $, we have 
\begin{equation}
	\!\!\!\!\!
	\frac{ \partial \mathfrak{R} \big(\mathbf{B}(\phi)\big) } { \partial\phi } = 
	\mathfrak{R} \left(  \frac{ \partial\mathbf{B}(\phi) } { \partial\phi } \right),\!
	\frac{ \partial \mathfrak{I} \big(\mathbf{B}(\phi)\big) } { \partial\phi } = 
	\mathfrak{I} \left(  \frac{ \partial\mathbf{B}(\phi) } { \partial\phi } \right). \!\!
\end{equation}

\subsection{Derivatives in 3D-JPVE Domain}\label{sec:Derivatives_loc}

We first define
\begin{gather}
	\mathbf{b}_{n,i} =  \mathbf{q}_i - \mathbf{p}_n,\; 
	s_{n,i} =  \frac{1} { \| \mathbf{b}_{n,i} \| },\; 
	\mathbf{k}_{n,i} = s_{n,i} \mathbf{b}_{n,i} ,\; i=1,2,
	\nonumber \\
	\!\!
    \mathbf{r}_{\ell} = [\mathbf{R}]_{:,\ell},\; \varpi_{n,\ell} = \mathbf{r}_\ell^{\mathsf{T}}\mathbf{b}_{n,2}, \;
	\varpi_{n,21} = \frac{\varpi_{n,2}}{\varpi_{n,1}},\;\ell = 1,2,3.\!\!
\end{gather}
Then the derivatives in 3D-JPVE domain can be written as
\begin{gather}
	\frac{ \partial d_{n,i} }{ \partial \mathbf{p} } = -\mathbf{k}_{n,i},\,
	\frac{ \partial d_{n,i} }{ \partial \mathbf{v} } = t_{n1}\frac{ \partial d_{n,i} }{ \partial \mathbf{p} },\,
	\frac{ \partial d_{n,i} }{ \partial B } = 1,\,
	\frac{ \partial d_{n,i} }{ \partial D } = t_{n1},
	\nonumber \\
	\frac{ \partial r_{n,i} }{ \partial \mathbf{p} } = 
	s^3_{n,i} \mathbf{b}_{n,i}^{\mathsf{T}} \mathbf{v} \mathbf{b}_{n,i} - s_{n,i}\mathbf{v}, \;
	\frac{ \partial r_{n,i} }{ \partial B } = 0,
	\nonumber \\
	\frac{ \partial r_{n,i} }{ \partial \mathbf{v} } = 
		s_{n,i} (\mathbf{b}_{n,i}-t_{n1}\mathbf{v}) + t_{n1}s^3_{n,i} \mathbf{b}_{n,i}^{\mathsf{T}} \mathbf{v} \mathbf{b}_{n,i}, \;
	\frac{ \partial r_{n,i} }{ \partial D } = 1,
\nonumber
\end{gather}
\begin{gather}
	\frac{ \partial \varphi_{n,\mathrm{az}} }{ \partial \mathbf{p} } = 
	\frac{1} { \varpi_{n,1}^2 (\varpi^2_{n,21} + 1) } 
	( \varpi_{n,2} \mathbf{r}_1 - \varpi_{n,1}\mathbf{r}_2 ),
	\nonumber \\
	\frac{ \partial \varphi_{n,\mathrm{az}} }{ \partial \mathbf{v} } = 
	t_{n1} \frac{ \partial \varphi_{n,\mathrm{az}} }{ \partial \mathbf{p} },\;
	\frac{ \partial \varphi_{n,\mathrm{az}} }{ \partial B } = 
	\frac{ \partial \varphi_{n,\mathrm{az}} }{ \partial D } = 0,
	\nonumber \\
	\frac{ \partial \varphi_{n,\mathrm{el}} }{ \partial \mathbf{p} } = 
	\frac{s_{n,2}} { \sqrt{ 1 - (s_{n,2}\varpi_{n,3})^2 } } 
	( s_{n,2}^2 \varpi_{n,3} \mathbf{b}_{n,2} - \mathbf{r}_3 ),
	\nonumber \\
	\frac{ \partial \varphi_{n,\mathrm{el}} }{ \partial \mathbf{v} } = 
	t_{n1} \frac{ \partial \varphi_{n,\mathrm{el}} }{ \partial \mathbf{p} },\;
	\frac{ \partial \varphi_{n,\mathrm{el}} }{ \partial B } = 
	\frac{ \partial \varphi_{n,\mathrm{el}} }{ \partial D } = 0.
	\label{eq:deriv_loc}
\end{gather}

\section{\texorpdfstring {Elements of $ \bm{\varepsilon} $ and $ \mathbf{J}_{\mathrm{err}} $}{Elements of $ \varepsilon $ and $ \mathbf{J}_{\mathrm{err}} $}} \label{sec:error}

Consider the estimation errors in $ \bm{\varepsilon} $ are stacked epoch-by-epoch. For $ n=1,\ldots,N $, we have
\begin{align}
	[\bm{\varepsilon}]_{4n-3} &= \Delta d_{n,12}^2 + 2d^{\circ}_{n,2}\mathbf{e}_n^{\mathsf{T}}\Delta \mathbf{e}_n\Delta d_{n,12}
	\nonumber \\
	& + 2d^{\circ}_{n,1}\Delta d_{n,12} + 2d^{\circ}_{n,2}d_{n,12}\mathbf{e}_n^{\mathsf{T}}\Delta \mathbf{e}_n,
	\nonumber \\
	[\bm{\varepsilon}]_{4n-2} &= \Delta d_{n,12} \Delta r_{n,12} + \mathbf{v}^{\mathsf{T}}\mathbf{R}\Delta\mathbf{e}_n\Delta d_{n,12} 
	\nonumber \\
	&+ d^{\circ}_{n,2}\mathbf{e}_n^{\mathsf{T}}\Delta\mathbf{e}_n\Delta r_{n,12}
	+ r_{n,1}^{\circ} \Delta d_{n,12} 
	\nonumber \\
	&+ d_{n,1}^{\circ} \Delta r_{n,12} 
	+ ( d_{n,2}^{\circ}r_{n,12}\mathbf{e}_n^{\mathsf{T}} + d_{n,12} \mathbf{v}^{\mathsf{T}} \mathbf{R}) \Delta\mathbf{e}_n,
	\nonumber \\
	[\bm{\varepsilon}]_{4n-1} &= d_{n,2}^{\circ} \mathbf{e}_n^{\mathsf{T}} \Delta \mathbf{f}_n,
	\nonumber \\
	[\bm{\varepsilon}]_{4n} &= d_{n,2}^{\circ} \mathbf{e}_n^{\mathsf{T}} \Delta \mathbf{g}_n,
\end{align}
and
\begin{gather}
	\!\![\mathbf{J}_{\mathrm{err}}]_{4n-3,4n-3} = 2 d^{\circ}_{n,1},
	\nonumber \\
	\!\![\mathbf{J}_{\mathrm{err}}]_{4n-2,4n-3} = r^{\circ}_{n,1},\;
	[\mathbf{J}_{\mathrm{err}}]_{4n-2,4n-2} = d^{\circ}_{n,1},
	\nonumber \\
	\!\![\mathbf{J}_{\mathrm{err}}]_{4n-2,4n-1} 
	= \mathbf{v}^{\mathsf{T}} \mathbf{R} d_{n,12} \dot{\mathbf{e}}_{\mathrm{az}}(\bm{\varphi}_n),
	\nonumber \\
	\!\![\mathbf{J}_{\mathrm{err}}]_{4n-2,4n} 
	= \mathbf{v}^{\mathsf{T}} \mathbf{R} d_{n,12} \dot{\mathbf{e}}_{\mathrm{el}}(\bm{\varphi}_n),
	\nonumber \\
	\!\!\!\!\!\![\mathbf{J}_{\mathrm{err}}]_{4n-1,4n-1}
	= - d^{\circ}_{n,2} \cos(\varphi_{n,\mathrm{el}}),\;
	[\mathbf{J}_{\mathrm{err}}]_{4n,4n}
	= - d^{\circ}_{n,2}, 
\end{gather}
where $ \dot{\mathbf{e}}_{\mathrm{az}}(\bm{\varphi}_n) $ and $ \dot{\mathbf{e}}_{\mathrm{el}}(\bm{\varphi}_n) $ are defined in \eqref{eq:deriv_defi}. The other elements of $ \mathbf{J}_{\mathrm{err}} $ are all zeros.

\bibliographystyle{IEEEtran}
\bibliography{IEEEabrv,ref.bib}

\end{document}